\documentclass[useAMS,usegraphicx,usenatbib]{mn2e}

\usepackage{txfonts}

\usepackage{color} 
\usepackage[normalem]{ulem} 

\title[Magnetic field and torque in accretion discs]{Magnetic
field structure and torque in accretion discs around millisecond pulsars}

\author[L. Naso, W. Klu\'zniak and J. C. Miller]
{L. Naso$^{1,2}$\thanks{E-mail: luca.naso@gmail.com}, 
W. Klu\'zniak$^{3}$ and 
J. C. Miller$^{4}$\\
$^{1}$National Astronomical Observatories of China, Chinese Academy of
Sciences, A20 Datun Road, Chaoyang District, Beijing 100012, China\\
$^{2}$Edisonweb, Via I. Silone 21, 95040, Mirabella Imbaccari, Italy\\
$^{3}$Nicolaus Copernicus Astronomical Center, Polish Academy of Sciences, 
ul. Bartycka, 18, PL-00-716 Warszawa, Poland\\
$^{4}$Department of Physics (Astrophysics), University of Oxford, Keble
Road, Oxford OX1 3RH, UK}

\begin{document}

\date{Accepted 2013 August 6. Received 2013 July 5; in original form 2012 October 29}

\pagerange{\pageref{firstpage}--\pageref{lastpage}} \pubyear{}

\maketitle

\label{firstpage}

\begin{abstract}
 Millisecond pulsars are rather weakly-magnetized neutron stars which are 
thought to have been spun up by disc accretion, with magnetic linkage between 
the star and the disc playing a key role. Their spin history depends 
sensitively on details of the magnetic field structure, but
idealized models from the 1980s and 1990s are still commonly used for 
calculating the magnetic field components. This paper is the third in a series 
presenting results from a step-by-step analysis which we are making of the 
problem, starting with very simple models and then progressively including 
additional features one at a time, with the aim of gaining new insights into 
the mechanisms involved. In our first two papers, the magnetic field structure 
in the disc was calculated for a standard Shakura/Sunyaev model, by solving 
the magnetic induction equation numerically in the stationary limit within the 
kinematic approximation; here we consider a more general velocity field in the 
disc, including backflow. We find that the profiles of the poloidal and 
toroidal components of the magnetic field are fairly similar in the two cases 
but that they can be very different from those in the models mentioned above, 
giving important consequences for the torque exerted on the central object. In 
particular we find that, contrary to what is usually thought, some regions of 
the disc outward of the co-rotation point (rotating more slowly than the 
neutron star) may nevertheless contribute to {\it spinning up} the neutron star 
on account of the detailed structure of the magnetic field in those parts of 
disc.
 \end{abstract}


%

\section{Introduction}\label{sec:INT}
%
Accretion discs around magnetized neutron stars can greatly influence the 
stellar magnetic field, and magnetic field deformations can in turn have an 
effect on the angular momentum exchange between the disc and the central 
object. Here we investigate these deformations and their effect on the neutron 
star spin. We focus mostly on the progenitors of millisecond pulsars, which are 
commonly believed to be descendants of normal neutron stars that have been spun 
up and recycled back as radio pulsars by acquiring angular momentum from their 
companion via an accretion disc during the low-mass X-ray binary phase 
\citep{Aal82, RS82}.

The importance of magnetic fields in accretion discs around neutron stars was 
recognized already in the early paper by \citet{PR72}. It was already clear 
then that the most relevant character in the play is the profile of the 
toroidal component of the magnetic field, $B_\phi$. However, it was only in 
\citet{GL79a} that a detailed model was developed. They pointed out that 
turbulent motion, reconnection and the Kelvin--Helmholtz instability allow the 
stellar field to penetrate the disc and hence to be affected by its velocity 
field\footnote{The failure of total screening of the magnetic field in 
accretion discs was then confirmed by the calculations of \citet{MS97}. They 
performed two-dimensional numerical simulations with the resistive 
magnetohydrodynamics 
equations, and studied the time evolution of the region between the inner edge 
of the disc and the magnetosphere.}. By assuming a dipole profile for the 
poloidal component of the magnetic field, they calculated a corresponding 
profile for $B_\phi$.

One of the main consequences of having a non-zero Lorentz force inside the disc 
is the creation of a magnetic torque, which can drastically influence the spin 
history of the central object and also change the shape of the disc. 
\citet{GL79b} used the solution obtained in \citet{GL79a} to compute the torque 
exerted on the neutron star by a Keplerian disc. However in \citet{W87} it was 
shown that the solution obtained by \citet{GL79a} violated the induction 
equation and a consistent analytic profile for the toroidal field was instead 
calculated, as a stationary solution of the induction equation in axisymmetry. 
In that model the poloidal component of the magnetic field was assumed to be a 
pure dipole; the disc was assumed to be Keplerian, to be in corotation with the 
neutron star at the upper and lower surfaces and not to have any motion along 
the poloidal direction; also, all horizontal derivatives were neglected with 
respect to vertical derivatives. At almost the same time, \citet{C87} developed 
a similar one-dimensional model where, in addition, a non-Keplerian profile for 
the disc was considered and where the horizontal terms in the induction 
equations were also included. Both authors agreed in writing the toroidal 
component of the magnetic field in the following form, in cylindrical 
coordinates $(\varpi, \phi, z)$:
 \begin{equation}
\label{eq:bp_an}
B_\phi = \gamma_{\rm a} \, (\Omega_{\rm disc} - \Omega_{\rm s}) \, B_z
\, \tau_{\rm d} \,  \, {\rm ,}
\end{equation}
 where $\gamma_{\rm a}$ is the amplification factor, $\Omega_{\rm disc}$ and 
$\Omega_{\rm s}$ are the disc and stellar angular velocities, respectively, and 
$\tau_{\rm d}$ is the dissipation time--scale. For a dipole $B_z$, 
equation~(\ref{eq:bp_an}) implies $B_\phi \propto \, \Delta \Omega / \varpi^3$, 
where $\Delta \Omega \equiv\Omega_{\rm disc} - \Omega_{\rm s}$. In these 
models, 
the amplification factor $\gamma_{\rm a}$ was taken to be a constant not much 
greater than unity (it depends on the steepness of the transition in $\Omega$,
from disc motion in the equatorial plane to corotation at the disc surface). The
precise profile of $\tau_{\rm d}$ depends on what is the dominant mechanism for
dissipating the magnetic field. \citet{W95} considered three different cases,
with $\tau_{\rm d}$ being dominated by the Alfv\'en velocity, turbulent
diffusion and magnetic reconnection, respectively.

Both \citet{W87} and \citet{C87} employed equation~(\ref{eq:bp_an}) with a 
dipole vertical field to calculate the magnetic torque. They found that the 
magnetic contribution to the torque has the same sign as $\Delta\Omega$, i.e. 
the regions of the disc inwards of the corotation point rotate faster than the 
star, push the magnetic field lines forward and try to speed the star up, while 
the regions outwards of the corotation point rotate more slowly, drag the field 
lines backwards and spin the star down. \citet{C92} relaxed the approximation 
of zero poloidal velocity and studied the magnetic field structure only in the 
inner region of the disc. His analysis confirmed the results of the previous 
models.

\citet{ER00} addressed the problem from a complementary viewpoint, considering 
the influence of a given stellar magnetic field on the structure of the 
accretion disc in terms of its height and surface density. They also considered 
the back reaction on the magnetic field, solving both the induction equation 
(in 2D) and the disc-diffusion equation (in 1+1D), using different time steps. 
They showed that, when $B_z$ is a dipole, equation~(\ref{eq:bp_an}) is valid 
only to within a factor of $0.5$ -- $1.5$ on average; this factor changes with 
radius and can be as small as $0.2$ for large radii.

\citet{AP00} carried out numerical calculations to look for steady-state 
axisymmetric configurations of a force-free magnetosphere corotating with the 
central star, treating the interaction with the disc by means of a simple model 
in which the disc velocity profile had only an azimuthal component (i.e. no 
accretion flow) and horizontal derivatives in the disc were neglected with 
respect to vertical ones. Doing this, they found field-line inflation occurring 
immediately outside the corotation radius, caused by interaction between the 
stellar dipole field and the disc, with the field lines spreading out and the 
local field strength becoming weaker. Calculating the resulting torque on the 
star in a way similar to \citet{GL79b}, they found that this can then be up to 
two orders of magnitude smaller than what is obtained if the vertical field is 
taken to be a dipole (depending on the boundary conditions and the size of the 
numerical domain).

This phenomenon of magnetic field-line inflation had previously been 
discussed by a number of other authors in the 1990s (see \citealp{LB94,
Lal95,BH96},). It arises due to initially dipolar field lines 
being dragged round by interaction with the differentially rotating disc 
matter; the field-line twisting produced by this leads to inflation of the 
magnetosphere structure in response to increased magnetic pressure coming from 
the resulting toroidal component of the field. There can be associated line 
opening, leading to loss of part of the magnetic linkage between the disc and 
the star and to a wind being driven out along the open field lines. This has 
been studied in detail by \cite{ZF09} in the context of classical TTauri stars.

In \citeauthor{papI} (\citeyear{papI, papII}, hereafter referred to as Paper I and Paper II, respectively),
we commenced a systematic step-by-step study of the how the original magnetic 
field of the neutron star is distorted by its interaction with the disc. The 
strategy was to start with very simple models and then progressively include 
additional features one at a time, with the aim of gaining new insights into 
the mechanisms involved. We started off by focusing entirely on the structure 
of the magnetic field inside the disc, imposing simple dipolar boundary 
conditions at the top of it and taking pure vacuum outside. Consideration of 
interaction with a non-vacuum magnetosphere (including the phenomena of 
magnetic-line inflation mentioned earlier) is reserved for a subsequent stage 
of the step-by-step programme. In papers I and II, the stationary induction 
equation was solved numerically in two dimensions throughout the interior of 
the disc, calculating all of the magnetic field components (i.e. the poloidal 
field was not forced to be dipolar there). As in earlier work, axisymmetry was 
assumed and the kinematic approximation was used (i.e. the velocity field was 
specified and then kept fixed during the calculation without including back 
reaction from the magnetic field). The velocity field was not taken to be 
purely azimuthal but also had a radial component, as given by the standard 
$\alpha$-disc model of \citeauthor{SS73} (\citeyear{SS73}, hereafter S\&S), 
and a theta component, 
introduced so as to match the upper and lower boundary conditions (imposed 
through the use of a coronal layer above and below the disc)\footnote{The 
radial velocity $v_r$ was the same as that of the $\alpha$-disc; $v_\theta$ was 
zero in the disc, while in the corona it smoothly changed to match the boundary 
conditions; and $\Omega$ was Keplerian in the main part of the disc, smoothly 
changing to corotation at the inner edge and in the corona.}. The results of 
this analysis demonstrated inward dragging of the poloidal field by the 
accreting matter and showed that the profile of the toroidal component of the 
magnetic field can deviate significantly from that given by 
equation~(\ref{eq:bp_an}), depending on the values of the magnetic Reynolds 
numbers.

In this paper, we follow the same methodology as in papers I and II but 
apply it now to the disc velocity profile calculated by \citeauthor{KK00} 
(\citeyear{KK00}, hereafter 
referred to as K\&K). In their paper, K\&K presented a self-consistent analytic 
solution for the general structure of the three-dimensional velocity field 
inside a stationary axisymmetric $\alpha$-disc, and we use this here as the 
basis for our discussion. We again calculate all of the components of the 
magnetic field inside the disc: as in papers I and II, we find significant 
deviations of the poloidal component away from a pure dipole and of the 
toroidal component away from equation~(\ref{eq:bp_an}) whereas, when comparing 
between the results for the S\&S and K\&K velocity prescriptions (which are 
very different), we find that the main overall behaviour is qualitatively 
rather similar.

We also compute the contributions to the net torque acting on the neutron 
star which would come from magnetic linkage with the disc. Instead of using a 
simplified expression for the torque, as in the early models, we calculate it 
numerically, within our model assumptions, by computing the moment of the 
Lorentz force using the magnetic field configurations obtained as above. This 
calculation will need to be revisited when we subsequently include improved 
boundary conditions representing a join on to a non-vacuum magnetosphere, but 
already we are seeing here striking results concerning which regions of the 
disc spin the star up or down.

In the next section, we describe our model in more detail. In 
Section~\ref{sec:EQU}, we discuss the equations used for our calculations: we 
write down the induction equation, describe the velocity field and derive the 
equation for the torque calculation. In Section~\ref{sec:MFS}, we present our 
results for the magnetic field structure, while those regarding the torque are 
presented in Section~\ref{sec:TOR}. In Section~\ref{sec:CMP}, we compare our 
results with those from papers I and II and finally, we summarize our analysis 
in Section~\ref{sec:CON}.

\section{Model}\label{sec:MOD}
%
 Our model consists of a neutron star with a dipole magnetic field which 
is aligned with the common rotation axis of the star and the (corotating) 
disc, everything being axisymmetric. As mentioned above, at this stage of the 
project, we are focusing entirely on the structure of the magnetic field inside 
the disc, imposing simple dipolar boundary conditions at the top of it and 
taking pure vacuum outside. The magnetic field is taken to be sufficiently 
weak in the main part of the disc, so that it does not produce any significant 
back reaction on the fluid flow. The approximation of neglecting this 
back reaction (the kinematic approximation) enormously simplifies the set of 
equations that one needs to solve in order to study the magnetic field 
structure. In fact, one then has to solve only the magnetic induction equation, 
which we do by looking for a stationary configuration of the magnetic field.

It is reasonable to suppose that close to the stellar surface the magnetic 
field is strong and dipolar: we are considering a surface field of $B \sim 3 
\times 10^8$~G, which is typical for millisecond pulsars (see e.g. 
\citealp{ZK06}). Because of the $1/r^3$ dipolar fall off, as one moves away from 
the star, the magnetic field intensity decreases outwards quite rapidly and the 
field lines become progressively more distorted by the matter in the disc. As 
shown in papers I and II, this deformation is twofold: (1) the angular rotation 
stretches any dipolar field line in the azimuthal direction, thus creating a 
toroidal magnetic-field component, and (2) the radial motion likewise has an 
effect by tending to pull any field line inwards with the accretion flow (or 
outwards in the case of a backflow).

Close to the stellar surface, the magnetic field is strong and completely 
dominates the dynamics of the matter, i.e. this is the opposite regime to the 
one in which the kinematic approximation holds. However, for exactly that 
reason, in this region we can safely assume that the magnetic field is close to 
dipolar. Our strategy is therefore not to make any detailed study of what 
happens in the close vicinity of the neutron star but instead to take the 
magnetic field there as just being a dipole. We place the inner edge of the 
disc at the Alfv\'en radius\footnote{The precise location of the inner edge of 
the disc is still a debated issue and several possibilities have been put 
forward, see e.g. appendix A in \citet{KR07}. The main proposals differ by no 
more than a factor of $\sim 2$ in the radius of the inner edge.} and solve the 
induction equation from there outwards. The innermost part of the disc is a 
transition region where the magnetic field does not completely dominate the 
dynamics of the matter, but where the magnetic torque is sufficient to alter 
the angular velocity of the matter flow away from Keplerian. We treat this 
region (which we refer to as the `inner disc') by imposing a suitable 
sub-Keplerian profile for the angular velocity there. The solution of the 
induction equation in this region is used only to provide suitable boundary 
conditions for the solution further out, where the kinematic approximation 
becomes reasonable.

There is another region where the kinematic approximation may not be suitable, 
and this is near to the surface of the disc. Although the magnetic field 
intensity decreases quite rapidly with $r$ and can be very small for large $r$, 
the matter pressure here is even weaker since it decreases sharply with 
$\theta$, because of the change in density from the disc to the exterior (here 
taken to be vacuum). We model this aspect by introducing a boundary layer, 
which we call the corona, at the top and bottom surfaces of the disc. In this 
corona, the weak magnetic field is less prone to follow the motion of the 
low--density 
matter and we reproduce this behaviour in the model by using a larger 
value of the turbulent magnetic diffusivity there, which helps in reducing 
distortions away from a dipole (i.e. away from the boundary conditions). We use 
this `by-hand-enhancement' of the diffusivity also in the inner part of the 
disc mentioned earlier.

Throughout this paper, we consider a canonical $1.4~M_\odot$ neutron star with 
radius $10$ km, rotation period $P_{\rm NS} = 10$~ms and a surface magnetic 
field of $3 \times 10^8$~G, corresponding to a magnetic dipole moment of $\mu = 
3 \times 10^{26}$~G cm$^3$. The mass accretion rate is taken to be $6.30 \times 
10^{16}$~g s$^{-1}$.

\begin{figure}
 \centering
 \includegraphics[width=0.40\textwidth]{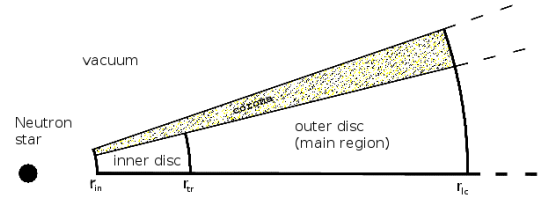}
 \caption{Schematic representation of our model (not to scale). We use $r_{\rm 
in} = 10\,r_{\rm g}$ and $r_{\rm tr} = 22 \, r_{\rm g}$. For a $10$~ms pulsar 
(with mass $1.4\,M_\odot$), $r_{\rm cor} = 18.8\,r_{\rm g}$ and $r_{\rm lc} = 
115\,r_{\rm g}$. The opening angles are $8^{\circ}$ for the disc alone and 
$10^{\circ}$ for the disc plus corona. The dashed lines indicate that the 
outer disc extends much further out than the main region shown here: the grid 
continues until $r_{\rm out} = 380 \, r_{\rm g}$ when we are solving for the 
toroidal component of the magnetic field and to $750 \, r_{\rm g}$ when we are 
solving for the poloidal one.}
 \label{fig:model}
\end{figure}

A sketch (not to scale) of the model is given in Fig.~\ref{fig:model}, where 
the two regions with larger diffusivity are clearly marked: they are the `inner 
disc' and the `corona'. The inner disc extends from the inner edge $r_{\rm in} 
= 10 \, r_{\rm g}$ to the transition radius $r_{\rm tr} = 22 \, r_{\rm g}$ 
(where $r_{\rm g} = 2GM/c^2 = 4.14\times 10^5$~cm is the Schwarzschild radius); 
the corona extends from $\theta = 80^\circ$ to $\theta = 82^\circ$ for all 
radii. The corotation radius is at $r_{\rm cor} = 18.8 \, r_{\rm g}$ and the 
light cylinder at $r_{\rm lc} = 115 \, r_{\rm g}$. To avoid strong dependence 
on the outer boundary conditions, our numerical domain of integration extends 
much further out than the light cylinder (dashed lines in the figure; more 
details are given in Sections~\ref{ssec:BCs} and \ref{ssec:SOL}).

%

\section{Equations}\label{sec:EQU}
%
 In this section, we derive the equations for calculating the magnetic field 
structure and comment on the profile of the velocity field and on the 
turbulent diffusivity as well as saying something further about the boundary 
conditions. We also derive equations for calculating the magnetic torque 
exerted by the disc on the neutron star and briefly comment on the solution 
method.

\subsection{Magnetic field}
 As mentioned previously, we are making our calculations using the kinematic 
approximation, within which the velocity field of the disc matter is specified 
separately (here following the K\&K solution) rather than being solved for 
together with the magnetic field calculation. Once the velocity field has been 
specified, the magnetic field structure can then be obtained by solving the 
induction equation:
 \begin{equation}
\partial_t \mathit{\mathbf{B}} = \nabla \times \left(  \mathbf{v} \times \mathbf{B} -
\eta \nabla \times \mathbf{B} \right) \,\, {\rm ,}
\label{eq:ind}
\end{equation}
 where $\mathbf{B}$ and $\mathbf{v}$ are the mean components of the magnetic 
field and velocity field respectively, and $\eta$ is the turbulent magnetic 
diffusivity. Resolving the full structure of turbulence and studying accretion 
driven by the magneto-rotational instability goes well beyond the scope of this 
paper. Interested readers are referred to \cite{Rom11} where the first global 
axisymmetric simulations of that kind are described. Here, we use average 
quantities and follow a phenomenological approach for the angular momentum 
transfer. As in papers I and II, we are neglecting the so-called 
$\alpha$-effect, which involves the appearance of a large--scale electromotive 
force because of the interaction between the turbulent fluctuations in the 
magnetic fields and the velocity fields. This term creates a growing solution 
for the magnetic field and is fundamental when one wants to study dynamo 
effects. In line with our approach, we want to use a simple model where one can 
unambiguously understand the roles played by the different contributing 
factors. Only afterwards will additional ingredients (such as dynamo action) be 
added and their effects studied by comparison with the previous analyses.

We use spherical coordinates ($r$, $\theta$, $\phi$), with the origin of the 
coordinate system placed at the centre of the neutron star and the unit vector 
$\mathbf{\hat z}$ pointing in the direction of the magnetic dipole axis (and 
the stellar rotation axis). In our model we are assuming axisymmetry 
($\partial_\phi[\dots] = 0$) and we want to find a stationary solution 
($\partial_t[\dots] = 0$). The three scalar components of 
equation~(\ref{eq:ind}) are then:
 \begin{eqnarray}
 0 &=& \partial_\theta \left\{ \sin\theta \left[ v_r B_\theta - v_\theta 
B_r - \frac{\eta}{r}[\partial_r(rB_\theta) - \partial_\theta B_r]\right] 
\right\} \label{eq:ind_r} \, \,{\rm ,} \\
 0 &=& \partial_r \left\{ r \hspace{0.5cm}\left[ v_r B_\theta - v_\theta 
B_r - \frac{\eta}{r}[\partial_r(rB_\theta) - \partial_\theta B_r] \right] 
\right\}\label{eq:ind_t} \, \,{\rm ,} \\
 0 &=& \partial_r \left\{ r \left[ v_\phi B_r - v_rB_\phi + \frac{\eta}{r} 
\partial_r(rB_\phi) \right] \right\} -\nonumber\\
&&\partial_\theta \left\{ v_\theta B_\phi - v_\phi B_\theta  -  
\frac{\eta}{r\sin\theta}\partial_\theta (B_\phi \sin\theta) \right\}  \,\, \rm{.}
\label{eq:ind_p}
\end{eqnarray}
 The first two equations are independent of $B_\phi$, and so we can solve these 
first for the poloidal components of the magnetic field and then solve the 
third equation afterwards for the toroidal component.

In order to solve for $B_r$ and $B_\theta$, one needs to specify profiles for 
$v_r$, $v_\theta$ and $\eta$ and to solve for $B_\phi$ one also needs a 
prescription for $v_\phi$. Then, given a set of consistent boundary conditions, 
the set of equations (\ref{eq:ind_r})--(\ref{eq:ind_p}) has a single unique 
solution.

\subsection{Velocity field}\label{ssec:VEL}
%
K\&K made a calculation of the three-dimensional structure for an 
$\alpha$-disc, taking the disc to be suitably thin and Taylor expanding the 
equations of viscous hydrodynamics in terms of a small parameter $\epsilon 
\equiv \mathcal{H} / \mathcal{R}$, where $\mathcal{H}$ is a characteristic 
vertical scaleheight in the disc and $\mathcal{R}$ is a characteristic radius 
(taken, in practice, to be the radius of the neutron star). The equation of 
state of the accreting medium was taken to be polytropic with constant specific 
entropy, implying that for any comoving fluid element, increases in entropy 
due to viscous dissipation are exactly balanced by losses due to radiation and 
conduction. They were able to find analytical solutions for the lowest order 
terms in their Taylor expansions and these demonstrated a flow structure which 
was entirely an inward accretion flow at the top and bottom of the disc but 
could have an outward backflow near to the mid--plane at radii beyond a certain 
value. A similar flow pattern, including equatorial zone backflow, was found by 
\cite{RG02} in a calculation including radiative transfer in an $\alpha$-disc 
modelled with a perfect gas equation of state. Clearly, this flow pattern is 
very different from that in standard S\&S discs and so it was of interest to 
repeat our earlier calculations of magnetic field distortions by S\&S discs 
also for these K\&K ones\footnote{Having this sort of stratified flow has 
sometimes been questioned on the grounds that the presence of large turbulent 
cells might prevent it. Here, we are using an $\alpha$-viscosity prescription 
with $\alpha \ll 1$, for which the issue of large turbulent cells need not 
arise. Indeed, having them would completely change the spectra and character of 
the S\&S solutions. More sophisticated MRI treatments for the effective 
viscosity have smaller-scale turbulence and {\it require} stratification for 
successful operation.}.

The general form of the Taylor expansions used by K\&K is given in section 2.1 
of their paper. We use here their prescriptions for the poloidal components of
the velocity field, $v_r$ and $v_z$, leaving the rotational velocity $\Omega$ as
being Keplerian in the main part of the disc. Due to parity considerations and
mass conservation, some of the terms appearing in the general form of the
expansions are zero, and one is left with the leading-order terms in the
expressions for $v_r$ and $v_z$ as being:
 \begin{eqnarray}
 \label{eq:KKexp_vr}
 && u = \frac{v_r}{\tilde{c}_s} = \epsilon \, u_1 + \mathcal{O}(\epsilon^3)  
\,\, {\rm ,} \\
 \label{eq:KKexp_vz}
 && v = \frac{v_z}{\tilde{c}_s} = \epsilon^2 \, v_2 + \mathcal{O}(\epsilon^4)  
\,\, {\rm ,}
\end{eqnarray}
 where $\tilde{c}_s$ is a characteristic sound speed in the disc. The analytic 
expressions for $u_1$ and $v_2$ are given by equations~2.53 and 2.56 of 
K\&K, respectively.

For implementing this velocity field within our calculation, we first choose 
the geometrical profile of the disc by specifying the disc height $h$ as a 
function of radius $r$ (note that here we are using $h$ and $r$ to represent 
physical lengths and not the scaled quantities used in K\&K). We choose this 
profile as being $h(r) = r \, \cos \theta_{\rm disc}$, where $\theta_{\rm disc}$
is the co-latitude of the disc surface (for which we take $\theta_{\rm disc} = 
82^{\circ}$). We sharply truncate the disc at an inner radius $r = r_{\rm in}$ 
(for which we take $r_{\rm in} = 10 \, r_{\rm g}$). We then specify a 
quantity which we call $v_0$, which is the value of the radial velocity $v_r$ 
at the point $(r_{\rm in},\theta_{\rm disc})$. Doing this turns out to be all
that is required in order to uniquely determine the velocity field in physical
units (i.e. as unscaled quantities). Changing the value of $v_0$ is equivalent
to changing the value of $\alpha$. From equation~2.55 of K\&K, one has
(neglecting $\mathcal{O}(\alpha^2)$ corrections): 
 \begin{equation}
 \label{eq:v0_alpha}
 v_0 \simeq 2 \, \alpha \, v_K \left( h/r \right)^2 \, \, \rm{,}
\end{equation}
 where $v_K$ is the Keplerian linear velocity and all quantities are calculated 
at $(r_{\rm in},\theta_{\rm disc})$.

\begin{figure}
 \begin{center}
\includegraphics[width=.45\textwidth]{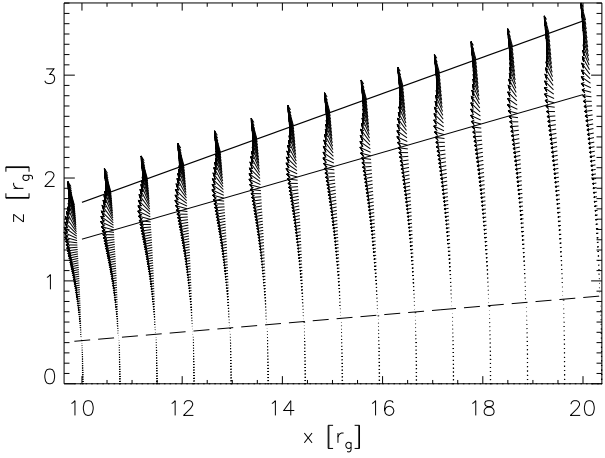}
 \end{center}
 \caption{Velocity-field vectors $v_r$ and $v_\theta$ plotted using Cartesian 
coordinates $x$ and $z$ measured in units of the Schwarzschild radius. (Note 
that we use these Cartesian coordinates for plots throughout this paper.) The 
dashed line separates the inflow region from the backflow one. The heavy solid 
line marks the boundary between the corona and the external vacuum; the other 
solid line shows the boundary between the disc and the corona.}
 \label{fig:pol_ref_vfield}
\end{figure}

\begin{figure}
 \begin{center}
\includegraphics[width=0.45\textwidth]{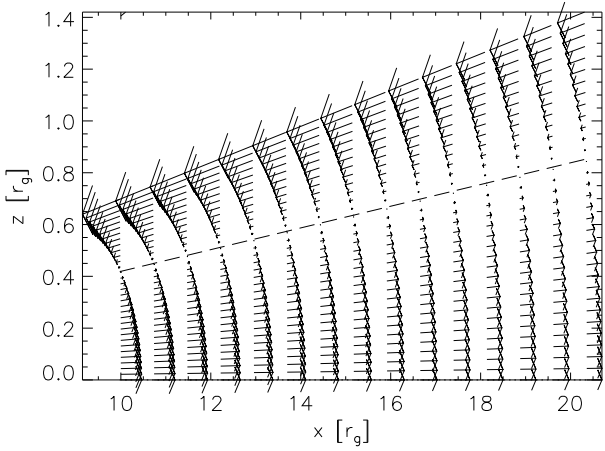}
 \end{center}
 \caption{Magnification of Fig.~\ref{fig:pol_ref_vfield} around the region 
where the flow changes direction. The region from $86^\circ$ to the equatorial 
plane is shown.}
 \label{fig:pol_ref_vfield_zoom}
\end{figure}

We follow the K\&K poloidal velocity profile through all of the disc and then 
modify it in the corona. There we first modify $v_\theta$ via an error function 
in order to match the boundary conditions smoothly and analytically. Then we 
modify $v_r$ in such a way that the total local kinetic energy remains 
constant. The resulting velocity-field vectors are shown in 
Figs~\ref{fig:pol_ref_vfield} and \ref{fig:pol_ref_vfield_zoom}. Note the 
appearance of a `stagnation surface' where both $v_r$ and $v_\theta$ go to 
zero, marking a change in direction of the flow. Having the surfaces where $v_r 
= 0$ and $v_\theta = 0$ coinciding is a particular property of the height 
profile that we are using here (with $h/r = {\rm constant}$).


As regards the angular rotation velocity, we follow the same profile as in 
Paper~II (see section 3.2 and fig. 2 of that paper). Here, we repeat the details 
again for completeness:
\begin{equation}
 \label{eq:omega}
 \Omega(r,\theta) = \left\{
 \begin{array}{cl}
  \Omega_K(r) & \hspace{0.3 cm} \mbox{in the main region:}\\
& \hspace{0.3 cm} \theta\in[\theta_{\rm disc},\pi/2], r\in[r_{\rm tr},
r_{lc}]\\
  & \\
  \mbox{smooth join in $\theta$} & \hspace{0.3 cm} \mbox{in the corona:}\\
 & \hspace{0.3 cm} \theta\in[\theta_{\rm cor}, \theta_{\rm disc}]\\
  & \\
  \mbox{smooth join in $r$} & \hspace{0.3 cm} \mbox{in the inner disc:}\\
 & \hspace{0.3 cm} r\in[r_{\rm in}, r_{\rm tr}]\\
  & \\
  \Omega_{\rm s} & \hspace{0.3 cm} \mbox{at the ghost zones: } \theta = 
\theta_{\rm cor} - \Delta \theta\\
& \hspace{0.3 cm} \mbox{at the inner edge: } r = r_{\rm in}
 \end{array}
 \right.
\end{equation}
 where $\theta_{\rm cor}$ is the upper surface of the corona, $\Delta \theta$ is
the
angular grid resolution, $\Omega_{\rm s}$ is the stellar spin rate, $\Omega_K$
is the Keplerian angular velocity and the two smooth connections are made using
the error functions given in equations~(\ref{eq:erf_t}) and (\ref{eq:erf_r})
below.

In the $\theta$, direction we write
 \begin{eqnarray}
\nonumber
 \tilde{\Omega}(r,\theta) &=& \Omega_K(r) \, f_1(\theta) + \Omega_{\rm s} \,
[1-f_1(\theta)]\\
 &=&[ \Omega_K(r) - \Omega_{\rm s}] \, f_1(\theta) + \Omega_{\rm s} \, \,{\rm ,}
\end{eqnarray}
where
\begin{equation}
\label{eq:erf_t}
 f_1(\theta) = \frac{1}{2}\left[ 1 + {\rm erf} \left(
\frac{\theta-\theta_{\rm disc}}{\delta_\theta} \right) \right] \, \,{\rm ,}
\end{equation}
with $\delta_\theta = 10^{-2}$ rad (i.e. $0.57^\circ$). Similarly, in the
radial direction:
 \begin{eqnarray}
 \nonumber
 \Omega(r,\theta) &=& \tilde{\Omega}(r,\theta)\,f_2(r)+\Omega_{\rm
s}\,[1-f_2(r)] \\
 &=&[ \tilde{\Omega}(r,\theta) - \Omega_{\rm s}] \, f_2(r) + \Omega_{\rm s} \, \,
{\rm ,}
\end{eqnarray}
where
\begin{equation}
\label{eq:erf_r}
 f_2(r) = \frac{1}{2}\left[ 1 + {\rm erf} \left( \frac{r-r_{\rm erf}}{\delta_r}
\right)
\right] \, \, {\rm ,}
\end{equation}
 with $\delta_r = 2 \, r_{\rm g}$ and $r_{\rm erf} = 14.4 \, r_{\rm g}$.
 
This angular velocity profile is Keplerian in all of the outer disc (see 
Fig.~\ref{fig:model}), while in the corona and in the inner disc it deviates 
from Keplerian in order to match corotation at the boundaries. On the 
equatorial plane, $\Omega$ starts to become sub-Keplerian at $r_1 = 16.67 
\, r_{\rm g}$ (where deviations are of the order of $1\%$) and reaches its 
maximum at $r = 15.18 \, r_{\rm g}$, after which 
it decreases towards corotation.

\subsection{Turbulent diffusivity}\label{ssec:ETA}
For the diffusivity, we follow the same approach as in Paper~II, writing
\begin{equation}
\label{eq:eta}
  \eta(r,\theta) = \eta_0 \, \left\{1 + \big[
\eta_\theta(\theta) + \eta_r(r) \big] \, \left[ \frac{\eta_{\rm c}}{\eta_0}
- 1 \right] \right\} \,\, {\rm ,}
\end{equation}
 where $\eta_0$ is the value in the main disc region and $\eta_{\rm c}$ is the 
value in the corona and inner disc (see Fig. \ref{fig:model}). We here take 
$\eta_0 = 10^{10}$~cm$^2$ s$^{-1}$ and $\eta_{\rm c} = 10^{12}$~cm$^2$ s$^{-1}$ 
as standard values. The magnitude of the diffusivity is often discussed in 
terms of the turbulent magnetic Prandtl number, which links it with that of the 
turbulent viscosity. There is a lot of uncertainty about which values to take 
and the one used here as our standard for $\eta_0$ may seem rather low. We have 
already discussed this issue in Paper~II and note also the discussions by 
\cite{RRS96} and others. We chose our value because it gives an interesting 
amount of field distortion in the disc, but it is clearly important to 
investigate the effects of varying it, and we also discuss this in the 
following sections.

For joining $\eta_\theta (\theta)$ and $\eta_r(r)$ between the low--$\eta$ and 
high--$\eta$ regions, we use joining functions of the form
 \begin{equation}
 f(x) = \frac{1}{2}\left[1 + {\rm erf}\left(\frac{- x + x_c} {d_x} \right) 
\right] \, \, {\rm ,}
\end{equation}
 where $x = \theta$ and $r$ for $\eta_\theta$ and $\eta_r$ respectively, with 
$x_c = \theta_{\rm disc}$ and $r_{\rm tr}$ in the two cases; $d_x$ is the width
of the transition in the error function, for which we use $d_r = 5\,r_{\rm g}$
in the radial direction and $d_\theta = 2^\circ$ in the theta direction.

\subsection{Boundary conditions}\label{ssec:BCs}
 We have already commented on the boundary conditions being used for the 
present stages of our calculations: vacuum everywhere outside of the disc (plus 
corona) and the neutron star, with a dipolar magnetic field throughout those 
vacuum regions. This is highly idealized, of course, but it is our working 
approximation for the moment. In order to have a smooth transition between the 
disc and the external vacuum, we include a boundary layer, a corona, and then 
impose boundary conditions at the top of it. We also impose dipole conditions 
at the inner and outer radial edges of the disc. At the inner edge, this is 
done because the magnetic field is strong there and so it is very hard to get 
any distortions there at all. What is done at the outer edge is irrelevant 
anyway because we are placing it so far away that it does not influence what is 
happening inside the light cylinder, and so we make the simple choice of 
imposing dipole conditions there as well. In Paper~I, we tested various kinds 
of boundary conditions (spherical lines, vertical lines and lines with 
$45^\circ$ inclination) and found that for poloidal magnetic field lines 
entering the extended disc (i.e. disc plus corona) at the same locations, their 
shape within the disc varied hardly at all among the different cases. However, 
the values for the {\it field strength} as a function of distance out along the 
surface of the extended disc will also vary depending on the structure of the 
external magnetic field and this will clearly influence the magnitude of the 
magnetic torque exerted by the disc.

We assume equatorial symmetry as well as axisymmetry and solve for only one 
quadrant. At our lower boundary in $\theta$ (i.e. at the equatorial plane), we 
require $B_\theta$ to be symmetric about the equatorial plane (i.e. 
$\partial_\theta B_\theta|_{\theta = \pi/2} = 0$), while $B_r$ and $B_\phi$ 
have antisymmetric conditions there (i.e. $B_r$ and $B_\phi$ are zero at this 
boundary, and at the first ghost grid-point beyond it they have values opposite 
to those at the last active grid-point before it)\footnote{We recall that 
boundary conditions for the poloidal component of the magnetic field are 
actually imposed on the magnetic stream function rather than on the magnetic 
field directly. See section $3$ of Paper~I for more details.}.

As regards the poloidal component of the velocity field, we follow K\&K exactly 
for the inner and lower boundaries. This means that on the equatorial plane 
$v_\theta = 0$ and $v_r$ is symmetric. This condition is equivalent to choosing 
a profile for $B_\theta$ at $\theta = \pi/2$ satisfying:
 \begin{equation}
 B_\theta(r, \pi/2) \,\, r \,\, v_r(r, \pi/2) = {\rm const} \,\,{\rm .}
\end{equation}
 At the top boundary, we have to modify the K\&K profile. In Paper~I (section 
3.2), we have already shown that if one requires the magnetic field to be a 
dipole, then the induction equation gives the following two conditions for the 
velocity field\footnote{Note that these conditions are valid only for a perfect 
dipole, i.e. $B_r = 2 \, k \cos\theta / r^3$, $B_\theta = k \sin\theta / r^3$ 
and $B_\phi = 0$ everywhere.}:
 \begin{eqnarray}
&& v_\theta =  (v_r /2) \tan\theta \,\\
&& \Omega \propto r^{-\gamma/2} \sin^\gamma \theta \,\, {\rm ,}
\end{eqnarray}
 where $\gamma$ can take any value. We use the first condition directly to 
calculate $v_\theta$ in the coronal layer and for $\Omega$ we use corotation, 
which is consistent with the second condition with $\gamma = 0$.

\subsection{Torque calculation}\label{sec:SUT}
 The magnitude of the torque exerted by the accreting matter on the neutron 
star, is clearly affected by the behaviour of the magnetic field in the 
magnetosphere. In the present simple model, we are taking vacuum above and 
below the extended disc so that the magnetic field there remains strictly 
dipolar, but in the general realistic case there can be field-line inflation, 
as mentioned earlier, which may also lead to breaking of some of the magnetic 
linkage between the disc and the star. It has been suggested that associated 
opening of field lines might give rise to jets and winds: see, for example, 
\cite{ZF13} concerning magnetospheric ejections. In this subsection, we 
consider the various contributions to the torque, first giving a rather general 
discussion and then specializing it to our simplified model.

\subsubsection{General discussion}

The disc can exert a torque on the central object via the link provided by 
magnetic field lines anchored at one end to the neutron star while the other 
end threads the disc, where the interaction with the moving matter creates a 
Lorentz force which can have a non-zero moment with respect to the central 
object. We call this contribution to the total torque $\mathbf{T}_{B-{\rm 
disc}}$.

As matter moves inwards, it encounters a progressively stronger magnetic field 
which eventually comes to completely dominate the motion of the plasma. This 
happens at the inner edge of the disc, where we suppose that matter leaves the 
disc following the magnetic field lines. If the magnetic field is then rigidly 
connected with the neutron star, it will force the matter to move with the same 
angular velocity as the central object. However, during this motion along the 
magnetic field lines, the matter is approaching the neutron star, and so 
although its angular velocity remains constant, there is a progressive decrease 
in its angular momentum (the cylindrical radial distance is decreasing). This 
occurs due to a torque exerted on it by the magnetic field lines, associated 
with which there is an equal and opposite torque tending to spin up the neutron 
star. We call this contribution $\mathbf{T}_{B-{\rm acc}}$.

Finally, when the accreting matter hits the neutron star surface, it transfers 
its residual angular momentum to the star. The torque concerned in this is 
typically very small, however, and we will neglect it here.

The total torque exerted on the star can then be written as: 
\begin{equation} 
\label{eq:3torques} 
\mathbf{T}_{\rm NS} = \mathbf{T}_{B-{\rm disc}} + \mathbf{T}_{B-{\rm acc}} \, \,
{\rm ,} 
\end{equation} 
with 
\begin{eqnarray}
\label{eq:T_Bd_def}
\mathbf{T}_{B-{\rm disc}} &=& \int\limits_{V} \mathbf{\tau}_{B-{\rm disc}} \, dV 
= - \int\limits_{V} \mathbf{r} \times \mathbf{F}_{\rm L} \, dV \, {\rm ,} \\
\label{eq:T_acc_def}
\mathbf{T}_{B-{\rm acc}} &=& \dot{m} \, \frac{2\pi}{P_{\rm NS}} \,
\left( \varpi_{\rm in}^2 - \varpi_{\rm surf}^2 \right)\,\mathbf{\hat{z}} \, \,
{\rm ,}
\end{eqnarray}
 where $\mathbf{\tau}_{B-{\rm disc}}$ is the moment of the Lorentz force 
($-\mathbf{F}_{\rm L}$) exerted on the neutron star by the local fluid element; 
$V$ is the volume containing all of the fluid elements associated with the 
disc which are magnetically linked to the neutron star and for which 
$\mathbf{F}_{\rm L}$ is non-zero; $\dot{m}$ is the accretion rate; $P_{\rm 
NS}$ is the neutron star spin period; $\varpi_{\rm in}$ is the distance 
from the rotation axis to the inner edge of the disc, while $\varpi_{\rm surf}$ 
is the corresponding distance for the point where the matter hits the surface 
of the neutron star and $\mathbf{\hat{z}}$ is the unit vector along the 
rotation axis. To the best of our knowledge, this is the first work where the 
magnetic torque is calculated by computing the integral of the Lorentz force 
through the whole disc, i.e.~by means of equation~(\ref{eq:T_Bd_def}). In the 
usual approach, only the value at the disc surface is considered.

Throughout this work we express torques with respect to the star, i.e. a 
positive value means that the torque is spinning the star up, while a negative 
one produces spin down.

The sign of $\mathbf{T}_{B-\rm acc}$ is always positive: it always acts so as 
to spin the star up (positive torque) because the angular momentum of the 
matter is decreasing when going from the inner edge of the disc to the neutron 
star surface. The sign of $\mathbf{T}_{B-{\rm disc}}$, however, is not obvious 
until the calculations are performed (see equation~\ref{eq:torque} below) and 
in general it can be either positive or negative.

We now consider in more detail the contribution to the torque due to the 
integrated moment of the Lorentz force (i.e. that coming from 
equation~\ref{eq:T_Bd_def}).

We are calculating the torque with respect to the origin of the coordinate 
system, and so the radial component of the force has zero moment. The force 
acting along the $\theta$ direction is always perpendicular to the position 
vector and so, if it is non-zero, it will always give a non-zero torque along 
the $\phi$ direction. However, since this component is antisymmetric with 
respect to the equatorial plane (while the position vector is symmetric), the 
integral of its moment will be zero. The only non-zero contribution to the net 
torque comes from the moment of the $\phi$ component. If we use $\mathbf{r}$ to 
indicate the position vector, and write $\mathbf{F}_L = (F_r \, 
\mathbf{\hat{r}}, F_\theta \, \mathbf{\hat{\theta}}, F_\phi \, 
\mathbf{\hat{\phi}} )$ with $\mathbf{\hat{r}}$, $\mathbf{\hat{\theta}}$ and 
$\mathbf{\hat{\phi}}$ being unit vectors in the $r$, $\theta$ and $\phi$ 
directions, we have
\begin{eqnarray}
\mathbf{\tau}_{B-\rm disc} &=&  -\mathbf{r} \times \left( F_r\mathbf{\hat{r}} + 
F_{\theta}\mathbf{\hat{\theta}} + F_\phi\mathbf{\hat{\phi}} \right) = 
-r\mathbf{\hat{r}} \times \left( F_{\theta}\mathbf{\hat{\theta}} + 
F_\phi\mathbf{\hat{\phi}} \right) \, \, {\rm ,} 
\\
\mathbf{\tau}_{B-\rm disc} &=& r \, F_\phi \, \mathbf{\hat{\theta}} 
- r \, F_{\theta} \, \mathbf{\hat{\phi}} \,\, {\rm ,} \\
\label{eq:T_Bdisc}
\mathbf{T}_{B-{\rm disc}} &=& \int\limits_{V} r \, F_\phi \, 
\mathbf{\hat{\theta}} \,dV =
- 2 \pi \iint\limits_{r\theta} r \, F_\phi \,\,\,r^2\sin\theta \,drd\theta 
\,\, \mathbf{\hat{z}} \,\, {\rm ,}
\end{eqnarray}
 where in the last equality we have changed the volume integral into three 
integrals over the three coordinate directions, integrated over $\phi$ (giving 
the $2\pi$ factor) and have explicitly written the direction of the torque as 
being along the vertical axis (this implies a change of sign because 
$\mathbf{\hat{\theta}}$ and $\mathbf{\hat{z}}$ have opposite directions).

From Maxwell's equations in cgs Gaussian units, we can write the general 
expression for the Lorentz force per unit volume as
\begin{equation}
 \mathbf{F_L} = \frac{1}{c}\mathbf{J} \times \mathbf{B} = \frac{\nabla \times 
\mathbf{B}}{4\pi}\times \mathbf{B} \,\, {\rm ,}
\end{equation}
whose $\phi$ component in spherical coordinates is
\begin{eqnarray}
\nonumber
 F_\phi & = & \frac{1}{4\pi}\frac{B_\theta}{r\,\sin\theta}\partial_\theta(B_\phi
\, \sin\theta) + \frac{1}{4\pi} \frac{B_r}{r}\partial_r(B_\phi\,r) \\
\label{eq:FL_phi}
&& - \frac{1}{4\pi}\frac{1}{r\sin\theta}\left(  B_\theta \partial_\phi B_\theta 
+ B_r \partial_\phi B_r \right) \,\, {\rm .}
\end{eqnarray}
 When considering axisymmetric models (as we are doing here), the third term can 
be neglected. The second term is also usually neglected for thin disc models 
because it involves derivatives with respect to the radius, while the first one 
involves vertical derivatives. The strategy of this work is to fully 
consider both radial and vertical structure and so we keep both terms in the 
torque calculation. Substituting equation~(\ref{eq:FL_phi}) into 
(\ref{eq:T_Bdisc}) one obtains
\begin{eqnarray}
\nonumber
\mathbf{T}_{B-{\rm disc}} =  \iint\limits_{r,\theta} \left\{ -\frac{1}{2} 
\left[  
B_\theta \partial_\theta(B_\phi\,\sin\theta) + \right.  \right. \\
\label{eq:torque}
\left. \vphantom{\frac{1}{2}} \left. B_r \partial_r(B_\phi\,r) \sin\theta 
\right]  
r^2 \right\} dr d\theta \,\, \mathbf{\hat{z}} \,\, {\rm .}
\end{eqnarray}
 We write the torque as the sum of two contributions in order to facilitate the
comparison with the thin disc models that neglect the radial term:
 \begin{equation}
\label{eq:torque_2terms}
\mathbf{T}_{B-{\rm disc}} = \mathbf{T}_{B-{\rm disc}}^{\theta} +  
\mathbf{T}_{B-{\rm disc}}^{r} \,\, {\rm ,}
\end{equation}
with
\begin{eqnarray}
\label{eq:T_Bdt}
\mathbf{T}_{B-{\rm disc}}^{\theta} &=& \iint\limits_{r,\theta} -\frac{1}{2} 
\left[ B_\theta \, \partial_\theta(B_\phi\,\sin\theta) \,r^2 \right]
 dr d\theta \,\, \mathbf{\hat{z}} \,\, {\rm ,}\\
\label{eq:T_Bdr}
\mathbf{T}_{B-{\rm disc}}^{r} &=&   \iint\limits_{r,\theta} -\frac{1}{2} \left[
B_r \partial_r(B_\phi\,r) \,r^2\sin\theta \,\right]
dr d\theta \,\, \mathbf{\hat{z}} \, \,{\rm .}
\end{eqnarray}

\subsubsection{Application to the model considered in this paper}
 As we have mentioned earlier (see Section~\ref{sec:MOD}), in the inner part of 
the disc our kinematic approximation breaks down because the magnetic pressure 
becomes non-negligible there with respect to the gas pressure (although it 
still remains smaller than the gas pressure until $r_{\rm in}$ is reached). In 
this part of the disc, we expect the angular velocity to become sub-Keplerian 
and finally to reach corotation with the neutron star at $r_{\rm in}$. Since we 
cannot reproduce this profile self-consistently (because we are using the 
kinematic approximation and are not solving an equation of motion), we change 
the $\Omega$ profile by hand in that region, as described in 
Section~\ref{ssec:VEL}. This should actually mimic the real situation rather 
well. For this region, instead of using equation~(\ref{eq:T_Bd_def}) to 
calculate the torque contribution, we instead calculate the back reaction on 
the field lines corresponding to the angular momentum loss by the matter; this 
would then be expected to be mainly transmitted to the neutron star (although 
it could also go into a jet). We calculate this torque contribution as
\begin{equation}
\label{eq:T_inner}
\mathbf{T}_\Omega = \dot{m} \left( \Omega_1 \,r_1^2 
- \Omega_{NS} r_{\rm in}^2 \right) \, \,\, \mathbf{\hat{z}} \,\, {\rm ,} 
\end{equation}
 where $r_1$ is the location on the equatorial plane where the angular velocity 
differs from the Keplerian value by $1\%$ ($r_1 = 16.67 \, r_{\rm g}$ with our 
model settings), and $\Omega_1$ is the vertically averaged value of the angular 
velocity at that radial location. More rigorously one should consider local 
values for $\dot{m}$ and $\Omega$ and integrate equation~(\ref{eq:T_inner}) 
over the vertical direction.\\

We then calculate the magnetic torque from the disc as the sum of two
components:
\begin{equation}
 \label{eq:TOmega+TB}
 \mathbf{T}_{B-{\rm disc}} = \mathbf{T}_\Omega + \mathbf{T}_B
\end{equation}
 where $\mathbf{T}_\Omega$ is given by equation~(\ref{eq:T_inner}) and 
$\mathbf{T}_B$ is given by equation~(\ref{eq:T_Bd_def}) integrated from $r_1$ 
out to the point where the contribution to the integral becomes negligible. We 
calculate $\mathbf{T}_{B-{\rm acc}}$ from equation~(\ref{eq:T_acc_def}), taking 
$\varpi_{\rm in}$ as being the distance from the rotation axis to the uppermost 
part of the inner edge of the disc (i.e. $r_{\rm in} \sin 80^\circ$) and 
getting $\varpi_{\rm surf}$ (the corresponding distance for the point where the 
infalling matter hits the surface of the neutron star) by assuming that the 
matter concerned follows dipole magnetic field lines after leaving the inner 
edge of the disc; this then gives $\varpi_{\rm surf} = r_{\rm NS} \, 
\sin(80^\circ)/2$.

\subsection{Solution method}\label{ssec:SOL}
 Equation~(\ref{eq:ind}) has been solved numerically using a Gauss--Seidel 
iterative scheme with an evenly spaced grid in $r$ and $\theta$. As in papers~I 
and~II, all variables and coordinates were written in a dimensionless form for 
the purposes of the calculation. Because of the partial decoupling between the 
poloidal and toroidal components of the magnetic field, we have first solved 
for $B_{\rm pol}$ (measured in units of $B_0 = 3 \times 10^8$~G), following the 
procedure described in Paper~I, and then for $B_\phi$ (measured in units of 
$B_\phi^0 = r_0^2 \, B_0$, where $r_0 = 2.5$ is the radius of the neutron star 
in units of $r_{\rm g}$) following the procedure described in Paper~II.

Details of the numerical scheme, including tests, have been given in papers I 
and II. From a technical point of view, the calculations performed here for 
obtaining the magnetic field structure are very similar to those of Paper~II, 
the only difference being in the functions used for the velocity components. We 
recall that we use a grid resolution of $\Delta r = 0.74\,r_{\rm g}$ and 
$\Delta\theta = 0.125^{\circ}$ for both $B_{\rm pol}$ and $B_\phi$. In both 
cases, we use a numerical domain which is much larger than the physical domain 
of interest in order to avoid dependence of the results on the outer boundary 
conditions.

In addition to solving for the magnetic field structure, in this paper we are 
also solving for the torque, which involves calculating the integral given in 
equation~(\ref{eq:torque}). We do this by using a trapezoidal scheme, as 
described in Section~\ref{sec:tor_disc}.

\section{Magnetic field structure}\label{sec:MFS}
We have studied a range of different models, so as to investigate the effect of
varying parameters, but first we present results for a representative fiducial
model to provide a standard against which to compare the others.

\subsection{Fiducial configuration}\label{sec:ref_conf} 
 Our fiducial run is characterized by intermediate values of the magnetic 
Reynolds numbers, that are defined as $R_{\rm m} \equiv l_c\,v_c / \eta_c$, 
where $l_c$, $v_c$ and $\eta_c$ are characteristic values for the length, 
velocity and diffusivity, respectively. For $l_c$, we take the radius of the inner edge of 
the disc (i.e. $l_c=10 \, r_{\rm g} = 4.136 \times 10^6$~cm); for $\eta_c$, 
we take $\eta_0$, as defined above; for $v_c$, we take either a representative 
radial velocity (for which we use $v_0$, as defined earlier) or a 
representative azimuthal velocity (for which we use the value of $v_\phi$ at 
the corotation radius, which we denote as $v_{0\,\phi}$). Depending on which of 
these velocities is used, we get either the poloidal or toroidal magnetic 
Reynolds number ($R_m^{\rm pol}$ or $R_m^{\rm tor}$, respectively). For our 
neutron star model, $v_{0\,\phi} = 4.88 \times 10^{9}$~cm~s$^{-1}$.

\begin{figure}
 \begin{center}
\includegraphics[width=0.45\textwidth]{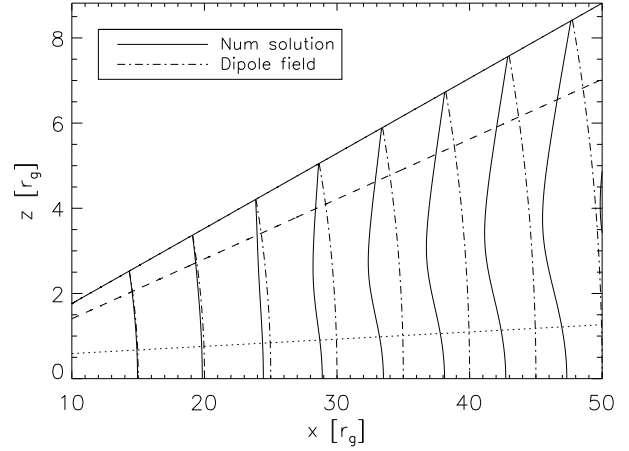}
 \caption{Poloidal field lines for the fiducial run. The dot--dashed lines are 
dipole field lines, shown for comparison; the dashed line is the boundary 
between the corona and the disc; the dotted line is the stagnation surface.}
 \label{fig:ref_Bpol}
 \end{center}
\end{figure}

\begin{figure}
 \begin{center}
\includegraphics[width=0.45\textwidth]{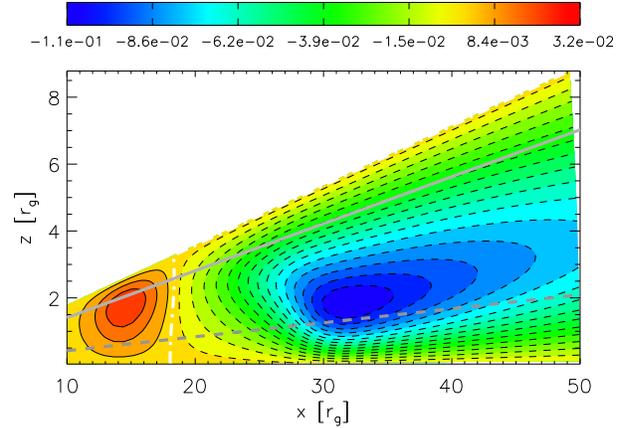}
 \end{center}
 \caption{$B_\phi$ contour plot for the fiducial run, in units of $B_\phi^0$. 
Solid contours indicate positive values, while dashed lines are used for the 
negative ones. The solid grey line is the boundary between the disc and the 
corona; the dashed grey line here marks the stagnation surface and a white 
dot--dashed line is used for the zero-level contour (which is just inwards of the 
corotation point).}
 \label{fig:ref_Bphi}
\end{figure}

The values that we use for our fiducial run are $R_m^{\rm pol} = 103$ and 
$R_m^{\rm tor} = 5 \times 10^5$, which we obtain by taking $v_0 = 10^6$~cm 
s$^{-1}$ and $\eta_0 = 4 \times 10^{10}$~cm$^2$~s$^{-1}$. The resulting 
structure of the poloidal component of the magnetic field $B_{\rm pol}$ is 
shown in Fig.~\ref{fig:ref_Bpol}. In Fig.~\ref{fig:ref_Bphi}, we show the 
contour plot of the toroidal component $B_\phi$. We find significant deviations 
of both $B_{\rm pol}$ and $B_\phi$ from the profiles which are usually 
considered, i.e. a dipole and the one given in equation~(\ref{eq:bp_an}), 
respectively.


The reason for the deviations of $B_{\rm pol}$ away from the dipole are clear: 
when field lines enter the disc, they are pushed inwards by the inward motion of 
the plasma and a negative $B_r$ appears. As one moves towards the equatorial 
plane, the radial component of the velocity field decreases, goes through the 
stagnation surface and then becomes positive in the region of backflow 
(as shown in Fig.~\ref{fig:pol_ref_vfield_zoom}). As a consequence, we see that 
the magnetic field lines are bent backwards, with $B_r$ going through a 
zero and then becoming positive again. In the inner disc region, the 
magnetic field follows the plasma motion to a lesser degree, as it should 
if one assumes that the magnetic stresses eventually truncate the disc (in fact 
$B_r$ always stays positive in this region). We recall that we obtain this 
effect because in this region we have a larger characteristic value for the 
turbulent diffusivity (giving a smaller local magnetic Reynolds number). 

\begin{figure}
 \begin{center}
\includegraphics[width=0.45\textwidth]{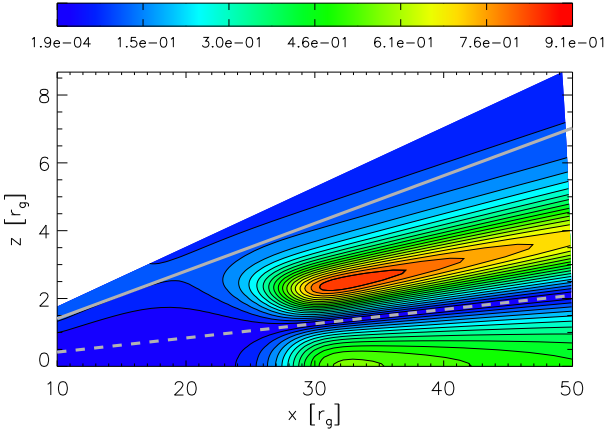}
 \end{center}
 \caption{Contour plot of $D_m^{\rm pol}$. Its peak is at about the same radial 
location as that of $D_m^{\rm tor}$ (see Fig.~\ref{fig:tor_ref_Dmcont}), and 
that of the second peak in the $B_\phi$ contour plots (see 
Figs.~\ref{fig:ref_Bphi}, \ref{fig:ref_B3cont}
and~\ref{fig:v0_0_eta0_1e9_B3cont}).}
 \label{fig:pol_ref_Dmcont}
\end{figure}

\begin{figure}
 \begin{center}
\includegraphics[width=0.45\textwidth]{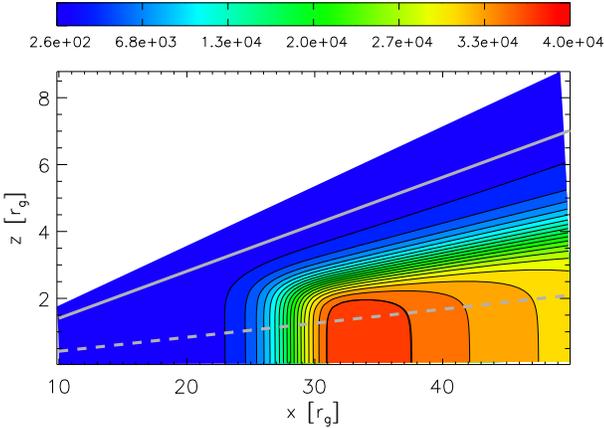}
 \end{center}
 \caption{Contour plot of $D_m^{\rm tor}$. Its peak is at about the same radial 
location as that of $D_m^{\rm pol}$ (see Fig.~\ref{fig:pol_ref_Dmcont}), and 
that of the second peak in the $B_\phi$ contour plots (see 
Figs.~\ref{fig:ref_Bphi}, \ref{fig:ref_B3cont}
and~\ref{fig:v0_0_eta0_1e9_B3cont}).}
 \label{fig:tor_ref_Dmcont}
\end{figure}

The toroidal component of the magnetic field has two extrema: the first being
positive and inwards of the corotation radius, and the second being negative and
outwards of it. This is exactly what one would expect if 
$B_\phi$ were to follow equation~(\ref{eq:bp_an}), with the vertical field 
being the same as for a dipole. However, if one analyses the profile in more 
detail one sees that while the first peak follows closely that of $B_\theta \, 
\Delta\Omega$, the second one appears at a much larger radius. A similar 
behaviour was also obtained in Paper~II and it was explained there by 
considering the magnetic distortion function $D_m$ (see section 5.2 of that 
paper). The distortion function was introduced in Paper~I: it is defined in the 
same way as the magnetic Reynolds number but, instead of taking characteristic 
values for the velocity and the diffusivity, it takes local values. Depending 
on which velocity component one considers, one can have a poloidal or toroidal 
magnetic distortion function:
\begin{eqnarray}
 D_{\rm m}^{\rm pol} &=& \frac{r_{\rm g}\sqrt{v_r^2 + v_\theta^2}}{\eta}\, ,\\
 D_{\rm m}^{\rm tor} &=& \frac{r_{\rm g}{v_\phi}}{\eta} \,\,{\rm .}
\end{eqnarray}
As in Paper~II we find that the position of the second peak in $B_\phi$ follows
the peak in the distortion functions (see Figs.~\ref{fig:pol_ref_Dmcont} and
\ref{fig:tor_ref_Dmcont}).


\begin{figure}
 \begin{center}
\includegraphics[width=0.41\textwidth]{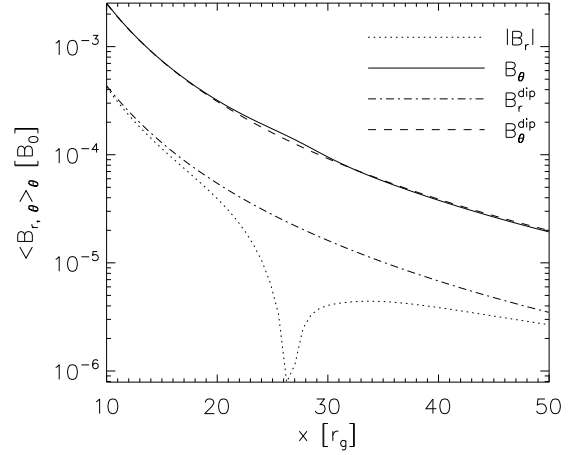}
 \end{center}
 \caption{Shell averages for the poloidal components of the magnetic field: 
$B_r$ (dotted) and $B_\theta$ (solid). $B_r$ changes sign at $x \sim 26$. A
dipole field is shown for comparison: its $r$ component is shown with the
dot--dashed line, while its $\theta$ component is shown with the dashed line.}
 \label{fig:pol_eta0_4e10_Bpolavg}
\end{figure}

In Fig.~\ref{fig:pol_eta0_4e10_Bpolavg}, we report the shell averages of the 
poloidal component of the magnetic field calculated in half of the disc, with 
the dipole ones for comparison. Distortions are present in both poloidal 
components, with the ones for $B_r$ being the more evident. This component even 
changes sign, at about $x \sim 26$ (the absolute value of the shell average is 
being plotted) and stays negative from that radius outwards, tending to zero 
from negative values. For the $\theta$ component (whose intensity is larger by 
about one order of magnitude), there is a weak but extended amplification in the 
inner part, between $x = 20$ and $x = 30$. This is the region where magnetic 
field lines accumulate (see Fig.~\ref{fig:ref_Bpol}). We have seen the same 
behaviour also in papers I and II and it is what one might expect in the 
transition region, where the magnetic field goes from being dominated by the 
plasma (large distortions) to dominating over the plasma (small distortion, 
dipole field).

\begin{figure}
 \begin{center}
\includegraphics[width=0.45\textwidth]{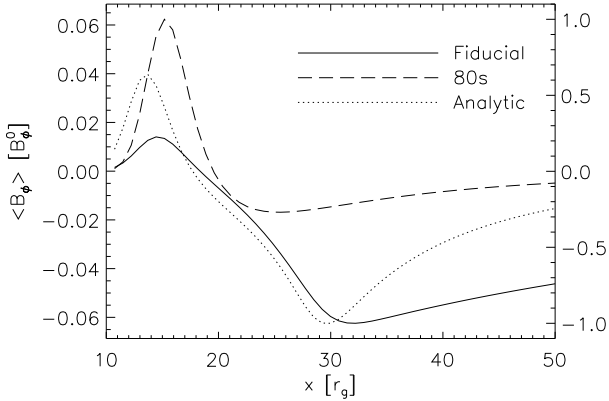}
\end{center}
 \caption{Shell averages for the toroidal component: the solid line is the 
result for our fiducial configuration; the dashed line is the prediction of 
equation~(\ref{eq:bp_an}) with a dipole $B_z$ and the dotted line is our
analytic prescription as obtained in Paper~II and reported in
equation~(\ref{eq:paperIIBp}). The scale for the fiducial shell average is
given on the left y-axis, while the other two curves are normalized to the
maximum of their absolute value (scale on the right). The 1980s result has its
zero for $B_\phi$ at corotation ($\sim 19\,r_{\rm g}$), whereas the fiducial
configuration has it at $\sim 17\,r_{\rm g}$ and the analytic profile has it at
$\sim 16\,r_{\rm g}$.}
 \label{fig:tor_eta0_4e10_B3cmp_avg}
\end{figure}

Fig.~\ref{fig:tor_eta0_4e10_B3cmp_avg} compares the $B_\phi$ shell average 
with the results of \citet{W87} and \citet{C87}\footnote{The curve for the 
models by Wang and by Campbell is calculated using equation~(\ref{eq:bp_an}), 
with a dipole field at $\theta = \pi/2$ being taken for $B_z$ and then 
normalizing to the maximum of the absolute value.} and includes also the 
analytic prescription suggested in Paper~II (see section 5.3 of that paper) that
we report here for clarity:
\begin{equation}
\label{eq:paperIIBp}
 B_\phi = \frac{r_g^2}{r\,\eta} \left[  \partial_r(r\, v_\phi\, B_r) +
\partial_\theta ( v_\phi \, B_\theta) \right] \,\,{\rm .}
\end{equation}
 If we apply this expression to the model considered in the present calculation, 
the agreement is not as good as it was in Paper~II. This is due to the breaking 
down of one of the assumptions made in deriving equation~(\ref{eq:paperIIBp}). 
In Paper~II, we assumed that the $B_\phi$ generation rate $\partial_t \, 
B_\phi|_+$ can be written as $[ \partial_r(r\, v_\phi\, B_r) + \partial_\theta 
( v_\phi \, B_\theta) ] / r$ and that the loss term $\partial_t \, B_\phi|_-\,$ 
can be written as $\,\eta B_\phi / r_{\rm g}^2$ and we then equated the two 
terms (because we were looking for a stationary solution). In the present model, 
there are some regions of the disc where the approximation for the loss term 
represents only a lower estimate. Nevertheless, in 
Fig.~\ref{fig:tor_eta0_4e10_B3cmp_avg}, one can see that the analytic expression 
still has a similar behaviour to that of the numerical solution and the 
agreement is quite good in the interpeak region, although it overestimates 
$B_\phi$ in the inner part and underestimates it in the outer one. All of the 
curves have the same general behaviour, with two peaks (the first positive and 
inwards of corotation, the second negative and outwards of corotation), and they 
all go through zero at approximately the same location (close to the corotation 
point). However, the relative amplitudes of the two extremal points are very 
different. The models of the 1980s, i.e. equation~(\ref{eq:bp_an}) with a 
dipole $B_z$, predict the first (positive) peak to be the strongest one, 
whereas our present calculations are showing just the opposite, with the second 
(negative) peak having an amplitude more than four times larger than the first 
one. Our analytic prescription can reproduce this behaviour, with the ratio 
between the amplitudes of the second and first peaks being larger than $1$, 
although that is still smaller (i.e. $\sim 1.6$) than what is required.

%

\subsection{Other configurations}\label{sec:CONF}
\subsubsection{List}
%
 In addition to the fiducial case, we analysed a range of other models to 
investigate the changes occurring in the magnetic field distortions when the 
parameters $v_0$ and $\eta_0$ (or $R_m^{\rm pol}$ and $R_m^{\rm tor}$) were 
varied over a wide range. Overall, we considered $13$ configurations, which we 
label `Mf' to denote Magnetic field structure. These are all listed in 
Table~\ref{tab:conf_mf}. For each configuration, we report the value of the 
turbulent magnetic diffusivity, $\eta_0$, the scale factor of the velocity 
field in the poloidal direction, $v_0$, the poloidal magnetic Reynolds number, 
$R_m^{\rm pol}$, the toroidal magnetic Reynolds number, $R_m^{\rm tor}$, and 
the value of $\alpha$ corresponding to $v_0$ (see 
equation~(\ref{eq:v0_alpha})).

\begin{table}
 \caption{Parameter values for the Mf configurations.}
\begin{center}
\begin{tabular}{cccccc}
\hline
Model & $\eta_0$ & $v_0$ & $R_m^{pol}$ & $R_m^{tor}$ & $\alpha$\\
& $(\rm{cm}^2 \, \rm{s}^{-1} )$ & $(\rm{cm} \, \rm{s}^{-1})$ &  &  &\\
\hline
Fiducial & $4 \times 10^{10}$ & $10^{6}$&$1.03\times 10^2$ & $5 \times 10^5$ &
$3.8 \times 10^{-3}$\\
\hline
Mf 1 & $10^{10}$ & $10^{6}$ & $4.14\times 10^2$ & $2 \times 10^6$ & $3.8 \times
10^{-3}$ \\
Mf 2 & $10^{11}$ & $10^{6}$ & $4.14\times 10^1$ & $2 \times 10^5$ & $3.8 \times
10^{-3}$ \\
Mf 3 & $10^{12}$ & $10^{6}$ & $4.14$ & $2 \times 10^4$ & $3.8 \times 10^{-3}$ 
\\
\hline
Mf 4 & $10^{10}$ & $10^{4}$ & $4.14$ & $2 \times 10^6$ & $3.8 \times 10^{-5}$ 
\\
Mf 5 & $10^{10}$ & $10^{5}$ & $4.14\times 10^1$ & $2 \times 10^6$ & $3.8 \times
10^{-4}$ \\
Mf 6 & $10^{10}$ & $2.5 \times 10^{5}$ & $1.03\times 10^2$ & $2 \times 10^6$ &
$9.4 \times 10^{-4}$ \\
\hline
Mf 7 & $10^{9{\phantom 0}} $ & $0$ & $0$ & $2 \times 10^7$ & $0$ \\
Mf 8& $10^{10}$ & $0$      & $0$ & $2 \times 10^6$ & $0$ \\
Mf 9& $10^{11}$ & $0$      & $0$ & $2 \times 10^5$ & $0$ \\
\hline
Mf 10& $10^{9{\phantom 0}} $ & $10^5$  & $4.14\times 10^2$ & $2 \times 10^6$ &
$3.8 \times 10^{-4}$ \\
\hline
\end{tabular}
 \label{tab:conf_mf}
\end{center}
\end{table}

Configurations Mf~1 -- Mf~3 differ from the fiducial case only in the value of 
$\eta_0$; the following three all have $\eta_0 = 10^{10}$~cm$^2$~s$^{-1}$ but 
different values of $v_0$; configurations Mf~7 -- Mf~9 all have $v_0 = 0$ and 
$\eta_0$ ranging from $10^9$ to $10^{11}$~cm$^2$~s$^{-1}$; 
Mf~10 differs from Mf~1 in the values of $\eta_0$, $v_0$ and also $v_{0\,\phi}$ 
(not reported in the table), but has the same magnetic Reynolds numbers as 
Mf~1. The characteristic velocity $v_{0\,\phi}$ is the same ($4.88 \times 
10^9$~cm~s$^{-1}$) for all of the configurations except for the last one, where 
we artificially lowered this value by a factor of $10$. In this case, 
therefore, the profile was no longer Keplerian. We were interested in this 
model because we wanted to test two cases which have different $\eta$ and $v$ 
(both poloidal and toroidal), but the same Reynolds numbers (Mf~1 and Mf~10).


\subsubsection{Results}\label{sec:other_conf}
\begin{figure}
 \begin{center}
\includegraphics[width=0.45\textwidth]{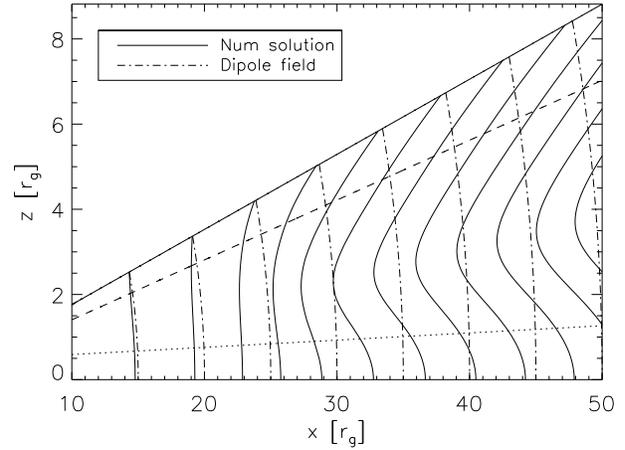}
 \end{center}
 \caption{Poloidal magnetic field lines for configurations with $R_m^{\rm pol} =
414$, i.e. Mf~1 and Mf~10.}
 \label{fig:Blines_eta0_1e10}
\end{figure}

\begin{figure}
 \begin{center}
\includegraphics[width=0.41\textwidth]{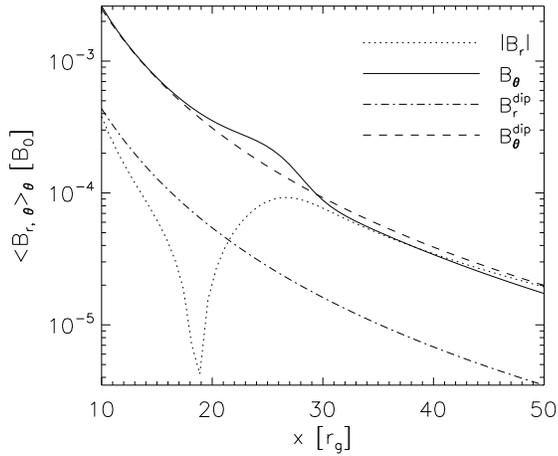} 
 \end{center}
 \caption{Shell average of the poloidal magnetic field components $B_r$ (dotted)
 and $B_\theta$ (solid) for configurations with $R_m^{\rm pol} = 414$, i.e. Mf~1
and Mf~10. $B_r$ changes sign at $x \sim 19$. Dipolar quantities are plotted for
comparison: the dot--dashed line is the shell average of the radial component,
the dashed line is that for the $\theta$ component.}
 \label{fig:avgpol_eta0_1e10}
\end{figure}

 In Paper~I we saw that what really matters for the poloidal field structure is 
the ratio between $v_0$ and $\eta_0$ (recall that in this paper $v_0$ is 
proportional to the disc $\alpha$, see Section~\ref{ssec:VEL}). This is 
confirmed in the present calculations and, in fact, configurations with the 
same $R_m^{\rm pol}$ (i.e. Mf~3 and Mf~4; Mf~1 and Fiducial; Mf~2 and Mf~5; 
Mf~7, Mf~8 and Mf~9) give identical $B_{\rm pol}$. When $R_m^{\rm pol}$ 
increases (i.e. when $v_0$ increases or $\eta_0$ decreases) the field gets 
progressively more frozen-in with the fluid and therefore distortions grow as 
well (see Fig.~\ref{fig:Blines_eta0_1e10}). As shown in 
Fig.~\ref{fig:avgpol_eta0_1e10}, this is clear also when comparing the shell 
averages. In the opposite regime instead (smaller $R_m^{\rm pol}$), the field 
tends to diffuse more, and deviations away from the dipole field are less 
evident (recall that we are imposing dipole boundary conditions). 
Configurations with high $R_m^{\rm pol}$ can lead to strong accumulation of 
magnetic field lines near to the inner edge of the disc. This magnetic field 
enhancement could favour jet-launching mechanisms, see e.g. \cite{Lii12}.

\begin{figure}
 \begin{center}
\includegraphics[width=0.45\textwidth]{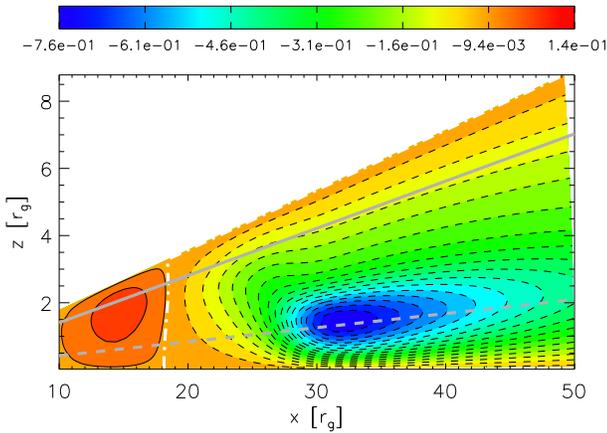}
 \end{center}
 \caption{Contour plot of $B_\phi$, in units of $B_\phi^0$. Positive levels are 
plotted with solid lines, negative levels with dashed lines. The solid grey 
line indicates the boundary between the disc and the corona; the dotted grey 
line marks the stagnation surface; while the white dot--dashed line is used for 
zero contour levels. This result is for configurations with $R_m^{\rm pol} = 
414$ and $R_m^{\rm tor} = 2 \times 10^6$, i.e. Mf~1 and Mf~10.}
 \label{fig:ref_B3cont}
\end{figure}

\begin{figure}
 \begin{center}
\includegraphics[width=0.45\textwidth]{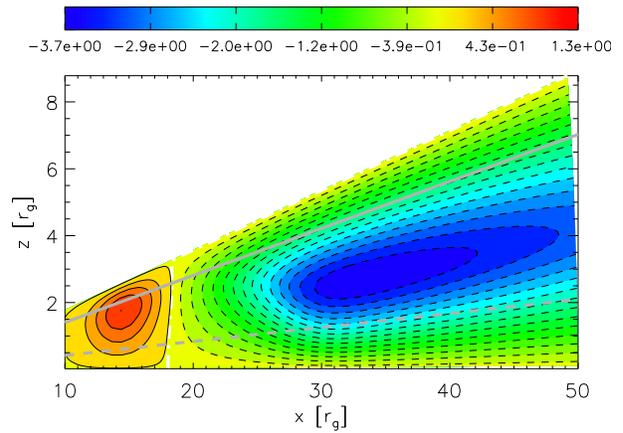}
 \end{center}
 \caption{As in Fig.~\ref{fig:ref_B3cont} but for configuration Mf~7, i.e. 
$R_m^{\rm pol} = 0$ and $R_m^{\rm tor} = 2 \times 10^7$. Note that the negative 
peak is here higher up in the disc than it was in Figs.~\ref{fig:ref_Bphi}
and~\ref{fig:ref_B3cont}.}
 \label{fig:v0_0_eta0_1e9_B3cont}
\end{figure}

Not surprisingly, this description does not suit the toroidal component as 
well, i.e. we see that configurations with the same ratio $v_0/\eta_0$ do not 
give the same $B_\phi$ values, but instead one should further consider 
$v_{0\,\phi}$. This is why we introduce the toroidal magnetic Reynolds number. 
In order to give the same profile for $B_\phi$ two configurations must have the 
same values of both $R_m^{\rm pol}$ and $R_m^{\rm tor}$. We show contour plots 
of $B_\phi$ for different pairs of the magnetic Reynolds numbers in 
Figs.~\ref{fig:ref_B3cont} and~\ref{fig:v0_0_eta0_1e9_B3cont}. The structure of 
the toroidal component of the magnetic field does not change very much with 
varying $R_m^{\rm pol}$, even for $R_m^{\rm pol} = 0$. When $R_m^{\rm pol}$ is 
held fixed and $R_m^{\rm tor}$ is varied, we find that $R_m^{\rm tor}$ affects 
only the scale factor for $B_\phi$ and that the shape of the contours is indeed 
determined only by $R_m^{\rm pol}$.


The behaviour described above is a consequence of the partial decoupling of the 
three components of the induction equation. As noted in papers~I and~II, under 
the assumptions of our models (i.e. axisymmetry, kinematic approximation and 
zero $\alpha$-effect\footnote{To avoid confusion with the $\alpha$ parameter of 
the S\&S $\alpha$-disc, let us clarify that the $\alpha$-effect we are 
referring to here is the one that we are neglecting in 
equation~(\ref{eq:ind}).}), the $r$ and $\theta$ components of the mean field 
induction equation do not depend on any $\phi$ quantity, and so we find that 
the $B_{\rm pol}$ structure depends only on $R_m^{\rm pol}$. The $\phi$ 
component instead depends on both toroidal and poloidal quantities, and so we 
find that $B_\phi$ depends on both $R_m^{\rm pol}$ and $R_m^{\rm tor}$.

As a final test for this description we ran a calculation for a case (Mf~10)
which has the same pair of Reynolds numbers as Mf~1, but given by different
values of the three quantities $\eta_0$, $v_0$ and $v_{0\,\phi}$. As expected,
we obtained identical results for the two cases.

\section{Torque exerted on the neutron star}\label{sec:TOR}
%
\subsection{Torque from the disc}\label{sec:tor_disc}
 In this paper, we have been concerned mainly with investigating the structure 
of the magnetic field within our extended disc, when simple dipolar boundary 
conditions are applied at the top and bottom of it, and we have seen features 
which are quite interesting (and which we will discuss further in the next 
section). Other additional elements still remain to be added to our model 
before it will be fully ready to be used for realistic application to spin up 
rates of millisecond pulsars. However, following our discussion in 
Section~\ref{sec:SUT}, it is useful to already apply the formulae there to 
calculate preliminary torque estimates for our present model, both from the 
point of view of providing a suitable benchmark against which to compare 
subsequent more complete results and also because this turns out to highlight 
some aspects which can have more general relevance.

The torque exerted by the disc on the star via the magnetic field is expressed 
by equations~(\ref{eq:torque}) and~(\ref{eq:T_inner}). The second of these can 
be calculated analytically, while the first one is calculated by using our 
numerical results for the magnetic field components. Along the radial 
direction, we integrate from where the angular velocity deviates from the 
Keplerian profile by $1\%$ (at $r_1 = 16.67 \, r_{\rm g}$) out to the point 
where the contribution to the integral becomes negligible, while along the 
$\theta$ direction we integrate over half the domain, from $80^\circ$ to 
$90^\circ$, and multiply by $2$ because of the symmetry. We use the same grid 
as that used for solving the induction equation (which is the grid where our 
solution for the magnetic field is defined) and use a trapezoidal scheme for 
two-dimensional quadratures, whose error is of the order of $O(\Delta r^2) + 
O(\Delta \theta^2)$. Given that $\Delta r = 0.74\,r_{\rm g}$, $\Delta\theta = 
2.2 \times 10^{-3}$ rad, and the integration range over $r$ and $\theta$ is 
$105\,r_g$ and $0.175$ rad respectively, we have a relative error of the 
order of $0.02\%$.

\subsubsection{List of configurations}
 We have calculated the torque for all of the configurations considered so far. 
However, that set of configurations was chosen for investigating the effect on 
magnetic field distortions of varying the parameters and is not ideally suited 
for studying in detail how the torque depends on the Reynolds numbers. Because 
of this, we define here 16 additional configurations, which are listed in 
Table~\ref{tab:conf_to} and are labelled `To' for Torque. As in 
Table~\ref{tab:conf_mf}, for each configuration, we report the values of the 
turbulent magnetic diffusivity, $\eta_0$, the representative velocity in the 
poloidal direction, $v_0$, the two magnetic Reynolds numbers, $R_m^{\rm pol}$ 
and $R_m^{\rm tor}$ and the parameter $\alpha$.

\begin{table}
 \caption{Parameter values for the configurations of the To group.}
\begin{center}
\begin{tabular}{cccccc}
\hline
Model & $\eta_0$ & $v_0$ & $R_m^{pol}$ & $R_m^{tor}$ & $\alpha$\\
& $(\rm{cm}^2 \, \rm{s}^{-1} )$ & $(\rm{cm} \, \rm{s}^{-1})$ &  &  &\\
\hline
To 1 & $10^{10}$ & $2.5   \times 10^2$ &$1.03\times10^{-1}$& $2 \times 10^6$ &
$9.4 \times 10^{-7}$ \\
To 2 & $10^{10}$ & $2.5   \times 10^3$ &            $1.03$ & $2 \times 10^6$ &
$9.4 \times 10^{-6}$ \\
To 3 & $10^{10}$ & $2.5   \times 10^4$ &$1.03\times10^{1}$ & $2 \times 10^6$ &
$9.4 \times 10^{-5}$ \\
To 4 & $10^{10}$ & $1.75  \times 10^5$ &  $7.24\times10^1$ & $2 \times 10^6$ &
$6.6 \times 10^{-4}$ \\
To 5 & $10^{10}$ & $5.0   \times 10^5$ & $2.07\times 10^2$ & $2 \times 10^6$ &
$1.9 \times 10^{-3}$ \\
To 6 & $10^{10}$ & $7.5   \times 10^5$ & $3.10\times 10^2$ & $2 \times 10^6$ &
$2.8 \times 10^{-3}$ \\
To 7 & $10^{10}$ & $1.5   \times 10^6$ & $6.20\times 10^2$ & $2 \times 10^6$ &
$5.7 \times 10^{-3}$ \\
To 8 & $10^{10}$ & $1.75  \times 10^6$ & $7.24\times 10^2$ & $2 \times 10^6$ &
$6.6 \times 10^{-3}$ \\
To 9 & $10^{10}$ & $1.87  \times 10^6$ & $7.76\times 10^2$ & $2 \times 10^6$ &
$7.1 \times 10^{-3}$ \\
To 10& $10^{10}$ & $2.0   \times 10^6$ & $8.27\times 10^2$ & $2 \times 10^6$ &
$7.6 \times 10^{-3}$ \\
To 11& $10^{10}$ & $2.25 \times 10^6$  & $9.31\times 10^2$ & $2 \times 10^6$ &
$8.5 \times 10^{-3}$ \\
To 12& $10^{10}$ & $3.75 \times 10^6$  & $1.55\times 10^3$ & $2 \times 10^6$ &
$1.4 \times 10^{-2}$ \\
\hline
To 13& $10^{11}$ & $5     \times 10^6$ & $2.07\times 10^2$ & $2 \times 10^5$ &
$1.9 \times 10^{-2}$\\
To 14& $10^{11}$ & $1.87  \times 10^7$ & $7.76\times 10^2$ & $2 \times 10^5$ &
$7.1 \times 10^{-2}$\\
To 15& $10^{11}$ & $4.5   \times 10^7$ & $1.86\times 10^3$ & $2 \times 10^5$ &
$1.7 \times 10^{-1}$\\
\hline
To 16& $10^{12}$ & $ 10^{7}$           & $4.14\times10^1$  & $2 \times 10^4$ &
$3.8 \times 10^{-2}$\\
\hline
\end{tabular}
 \label{tab:conf_to}
\end{center}
\end{table}

Configurations To~1 -- To~12 all refer to cases with $R_m^{\rm tor} = 2 \times 
10^6$ and have $R_m^{\rm pol}$ varying between $\sim 10^{-1}$ and $\sim 1.5 
\times 10^3$. In order to keep the toroidal Reynolds number constant, we have 
to keep $\eta_0$ constant as well and get the change in the poloidal Reynolds 
number by changing only $v_0$. Configurations To~13 -- To~15 have $R_m^{\rm 
tor} = 2 \times 10^5$ while the last one has $R_m^{\rm tor} = 2 \times 10^4$.

\subsubsection{Results}
 The torque contributions from the inner part of the disc and from the 
accreting matter inwards of $r_{\rm in}$ are the same for all of our
configurations, and can be calculated from equations~(\ref{eq:T_inner}) and 
(\ref{eq:T_acc_def}). Inserting the characteristic values being used in this 
work, these give
\begin{equation}
 \mathbf{T}_\Omega = 1.379 \times 10^{33} \,\,\mathbf{\hat{z}}  \, 
[\rm{dyn \, cm}] \,
\end{equation}
and
\begin{equation}
\label{eq:T_acc}
  \mathbf{T}_{B-\rm acc} = 0.646 \times 10^{33} \,\,\mathbf{\hat{z}} \, \, 
[\rm{dyn \,\, cm}] \, {\rm .}
\end{equation}
 Note that both of these are positive (i.e. acting in the direction of 
tending to give spin-up of the neutron star). The torque from the rest of the 
disc, $\mathbf{T}_B$, is calculated as described above and can, in principle, 
be either negative or positive (although it is usually negative in practice). 
The total torque exerted on the neutron star, $T_{\rm NS}$, is then the sum of 
these three contributions.

A summary of results for the Mf and To configurations is shown in 
Tables~\ref{tab:torque_mf} and~\ref{tab:torque_to}, respectively. (We drop the 
vector notation for $\mathbf{T}$ from here on.) In the second and third 
columns, we give the contribution from the outer part of the disc, $T_B$ and 
the radial contribution to it, $T_B^r$ (see equation~\ref{eq:torque_2terms}). 
Column four gives the total magnetic contribution from the disc 
($T_{B-\rm{disc}} = T_B + T_\Omega$). The fifth column gives the total torque 
acting on the neutron star, i.e. $T_{\rm NS} = T_{B-\rm{disc}} + T_{B - {\rm 
acc}}$. Note that the term $T_{B}^{r}$, which is usually neglected in the thin 
disc approximation, is actually quite large here and in many cases acts to 
oppose $T_{B}^{\theta}$. Including $T_{B}^{r}$ in two-dimensional calculations 
of the magnetic torque is therefore very important. $T_{B}^{\theta}$ remains 
dominant for all of these configurations, however, except for those with the 
highest values of $R_m^{\rm pol}$ (giving the largest radial distortions).

\begin{table}
 \caption{Values of the computed torque for the Mf configurations. $T_B$ is the 
torque due to the interaction between the magnetic field and the disc in the 
region outwards of $r_1$; $T_B^r$ is its radial component; $T_{B-{\rm disc}}$ is 
the magnetic contribution from all of the disc; $T_{\rm NS}$ is the total 
torque exerted on the neutron star. The torques are expressed in units of 
$10^{33}$~dyn~cm, with positive values indicating that they are acting in the 
direction of spinning up the neutron star while negative values give 
spin-down.}
 \begin{center}
\begin{tabular}{crrrrrr}
\hline
Model & $T_B$  & $T_B^r$ & $T_{B-{\rm disc}}$ & $T_{\rm NS}$ \\
\hline
Fiducial&$-1.61$& $0.34$ & $-0.23$ & $0.41$    \\
\hline
Mf 1 & $-6.68$  & $-5.12$& $-5.30$ & $-4.66$  \\
Mf 2 & $-0.47$  & $0.12$ & $0.91$  & $1.56$  \\
Mf 3 & $-0.03$  & $0.01$ & $1.35$  & $1.99$  \\
\hline
Mf 4 & $-3.22$  & $0.69$ & $-1.84$ & $-1.19$ \\
Mf 5 & $-4.66$  & $1.16$ & $-3.28$ & $-2.64$ \\
Mf 6 & $-6.44$  & $1.37$ & $-5.07$ & $-4.42$ \\
\hline
Mf 7 & $-30.4$  & $6.27$ & $-29.0$& $-28.35$ \\
Mf 8& $-3.04$  & $0.63$ & $-1.66$ & $-1.01$ \\
Mf 9& $-0.30$  & $0.06$ & $1.08$  & $1.72$ \\
\hline
Mf 10& $-6.68$  & $-5.12$& $-5.30$ & $-4.66$ \\
\hline
\end{tabular}
 \label{tab:torque_mf}
\end{center}
\end{table}
\begin{table}
 \caption{As for Table~\ref{tab:torque_mf} but for the To configurations.}
\begin{center}
\begin{tabular}{crrrrr}
\hline
Model & $T_B$  & $T_B^r$ & $T_{B-{\rm disc}}$ & $T_{\rm NS}$ \\
\hline
To 1 & $-3.04$ & $ 7.41$ & $-1.66$ & $-1.02$ \\
To 2 & $-3.08$ & $ 7.37$ & $-1.70$ & $-1.06$ \\
To 3 & $-3.48$ & $ 6.88$ & $-2.10$ & $-1.45$ \\
To 4 & $-5.65$ & $ 4.04$ & $-4.27$ & $-3.62$ \\
To 5 & $-7.88$ & $-0.16$ & $-6.50$ & $-5.86$ \\
To 6 & $-7.76$ & $-1.73$ & $-6.38$ & $-5.74$ \\
To 7 & $-3.57$ & $-1.49$ & $-2.19$ & $-1.54$ \\
To 8 & $-2.05$ & $-0.72$ & $-0.67$ & $-0.02$ \\
To 9 & $-1.35$ & $-0.30$ & $ 0.03$ & $ 0.67$ \\
To 10& $-0.71$ & $ 0.14$ & $ 0.67$ & $ 1.32$ \\
To 11& $ 0.41$ & $ 1.00$ & $ 1.79$ & $ 2.44$ \\
To 12& $ 2.53$ & $ 3.52$ & $ 3.91$ & $ 4.56$ \\
\hline
To 13& $-0.79$ & $ 0.01$ & $ 0.59$ & $ 1.24$ \\
To 14& $-0.14$ & $-0.03$ & $ 1.24$ & $ 1.89$ \\
To 15& $ 0.14$ & $ 0.26$ & $ 1.52$ & $ 2.17$ \\
\hline
To 16& $-0.05$ & $ 0.06$ & $ 1.33$ & $ 1.98$ \\
\hline
\end{tabular}
 \label{tab:torque_to}
\end{center}
\end{table}

In the usual description of the interaction between the magnetosphere and the 
disc, the sign of the torque caused by the magnetic field deformation is taken 
to be the same as that of $B_\phi$. However, this result follows after a 
succession of approximations have been made. In equation~(\ref{eq:torque}), 
which defines the magnetic torque, one can see that $B_\phi$ enters only via 
derivative terms. From the contour plots of $B_\phi$ (see 
Figs.~\ref{fig:ref_Bphi}, \ref{fig:ref_B3cont} 
and~\ref{fig:v0_0_eta0_1e9_B3cont}) it is clear that the sign of these does not 
always coincide with the sign of $B_\phi$ itself. This is why our results can 
be significantly different from those of the models developed in the 1980s (and 
subsequent calculations using a similar approach) and why we see changes in the 
sign of the magnetic torque without any change in the sign of $B_\phi$, but 
with only a modification of its two-dimensional structure (actually changes 
related to the poloidal component are larger here than those coming from the 
toroidal one). This aspect is considered further in Section~\ref{sec:spinUD} 
below, where we focus in particular on the question of which regions of the 
disc tend to spin the star up or down.

\subsection{Dependence on the magnetic Reynolds numbers}\label{sec:torque_Rm}
 Usually differences in the magnetic torque are caused by differences in 
the spin period of the star or in its magnetic field, but we have shown here 
that serious differences in the torque could also be caused by variations in 
the velocity field and the turbulent diffusivity, i.e. in the magnetic Reynolds 
numbers. We now want to describe better how this works.

We have already commented in Section~\ref{sec:other_conf} that $R_m^{\rm pol}$ 
fully determines the poloidal component of the magnetic field and $R_m^{\rm 
tor}$ enters as a scale factor in the toroidal component. Since the equation 
for calculating the magnetic torque is homogeneous in all of the magnetic field 
components (see equation~\ref{eq:torque}), it was to be expected that 
$R_m^{\rm tor}$ would affect the torque only as a scale factor. The other 
Reynolds number, $R_m^{\rm pol}$, instead influences the shape of the magnetic 
field lines and it is not trivial to predict its effect on the torque. In order 
to analyse this, we consider configurations with the same $R_m^{\rm tor}$ and 
different $R_m^{\rm pol}$ and we then plot the torque against $R_m^{\rm pol}$.

\begin{figure}
 \begin{center}
  \includegraphics[width=0.38\textwidth]{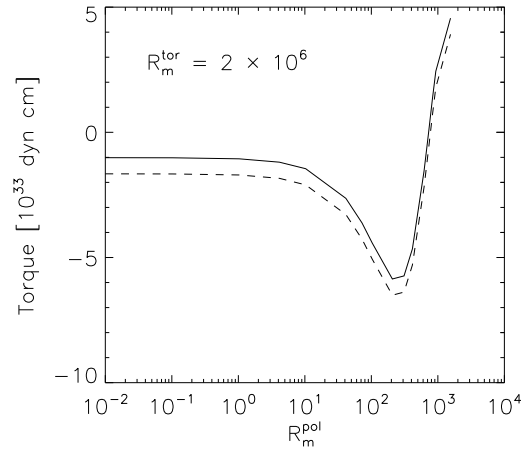}
 \end{center}
 \caption{Dependence of the torque on the poloidal magnetic Reynolds number 
$R_m^{\rm pol}$. Here, $R_m^{\rm tor}$ is fixed at $2 \times 10^6$. The solid 
line shows the total torque exerted on the neutron star, while the dashed line 
is for $T_{B-\rm{disc}}$ alone. The leftmost point is for $R_m^{\rm pol} = 0$ 
but is actually drawn at $10^{-2}$ so as to give a better visualization.}
 \label{fig:torque_Rmpol}
\end{figure}

Results are shown in Fig.~\ref{fig:torque_Rmpol}, where the poloidal magnetic 
Reynolds number ranges from $0$ to $1.55 \times 10^3$ and the toroidal one is 
kept fixed at $2 \times 10^6$. In Fig.~\ref{fig:torque_Rmpol}, we show both the 
total torque exerted on the neutron star and also the contribution coming from 
just the magnetic linkage with the disc, $T_{\rm B-disc}$. We have made similar 
calculations also for a different value of the toroidal magnetic Reynolds 
number, $R_m^{\rm tor} = 2 \times 10^5$ and found values for $T_{\rm B-disc}$ 
exactly a factor of 10 smaller, as expected.

As one can see from Fig.~\ref{fig:torque_Rmpol}, for very low $R_m^{\rm pol}$ 
the torque tends asymptotically to about $-1 \times 10^{33}$~dyn~cm. This means 
that when $B_r$ and $B_\theta$ are not distorted at all, i.e. the poloidal 
field is exactly a dipole, the disc still gives a magnetic contribution to the 
torque. This was of course expected, since one still has distortions in the 
$B_\phi$ component. Indeed, in most of the models in the literature, the 
poloidal field is assumed to be a dipole and yet the torque is non-zero. As 
deviations away from a dipole field increase (i.e. as $R_m^{\rm pol}$ 
increases), the torque becomes progressively more negative and reaches a 
minimum of $\sim -6\times10^{33}$~dyn~cm at $R_m^{\rm pol} \sim 200$. For even 
larger values of $R_m^{\rm pol}$, the torque starts to rise again and becomes 
positive.

Because of the proportionality between $R_m^{\rm tor}$ and $T_B$, we expect 
that as we lower $R_m^{\rm tor}$ the absolute value of $T_B$ will become 
progressively smaller. In particular, when $R_m^{\rm tor}$ reaches some 
critical value (say $R_m^{\rm zero}$) the negative peak in $T_B$ will be just 
enough to balance $T_\Omega + T_{B-{\rm acc}}$. In this case the total torque 
acting on the neutron star would always be positive except for the $R_m^{\rm 
pol}$ which gives the maximum torque, when it would instead be zero. When 
$R_m^{\rm tor}$ becomes smaller than a threshold value (say $R_m^{\rm const}$), 
$T_B$ would eventually become completely negligible with respect to $T_\Omega + 
T_{B-{\rm acc}}$ ($< 1\%$), so that the total torque $T_{\rm NS}$ would be of the
order of $2 \times 10^{33}$~dyn~cm for any value of $R_m^{\rm pol}$ or 
$\alpha$. By assuming that the linear scaling between $B_\phi$ and $R_m^{\rm 
tor}$ is valid for any $R_m^{\rm tor}$, we estimate these threshold values to 
be about $R_m^{\rm zero} = 4 \times 10^5$ and $R_m^{\rm const} = 4 \times 
10^4$.

We should make a careful caveat here, however. This discussion has been made 
within the context of our present simplified model. Including the effects of 
line inflation would be expected to decrease the magnitude of $T_B$, making 
spin-up more likely.

\subsection{Spin-up and spin-down regions}\label{sec:spinUD}
 Finally, we would like to comment in detail on the regions of the disc that 
spin the star up or down. In previous work, there has been a widespread 
consensus that the part of the disc which is inwards of the corotation radius 
gives a spin-up contribution, while that which is outwards of the corotation 
radius gives a negative contribution.

In order to investigate this aspect, we have calculated the contribution to the 
magnetic torque at every location inside the disc (i.e. the local value of the 
integrand in equation~(\ref{eq:torque})) and have then produced a contour plot 
of this magnetic torque density. The regions where it is negative act in the 
sense of spinning the star down, while those where it is positive act in the 
sense of spinning the star up. We have calculated this quantity for four key 
configurations, which have been chosen considering the behaviour shown in 
Fig.~\ref{fig:torque_Rmpol}. They are: (i) configuration Mf~8 with zero 
$R_m^{\rm pol}$, (ii) configuration To~5 with the largest negative torque, 
(iii) configuration To~8 with roughly zero net torque and (iv) configuration 
To~12 with the largest positive torque. We have considered configurations with 
the same value of $R_m^{\rm tor} = 2 \times 10^6$ (since the torque scales with 
$R_m^{\rm tor}$ comparisons should be made keeping this quantity constant) and 
with $R_m^{\rm pol} < 2 \times 10^3$ (for larger values distortions are too 
large and could be non-realistic).

\begin{figure}
 \begin{center}
\includegraphics[width=0.45\textwidth]{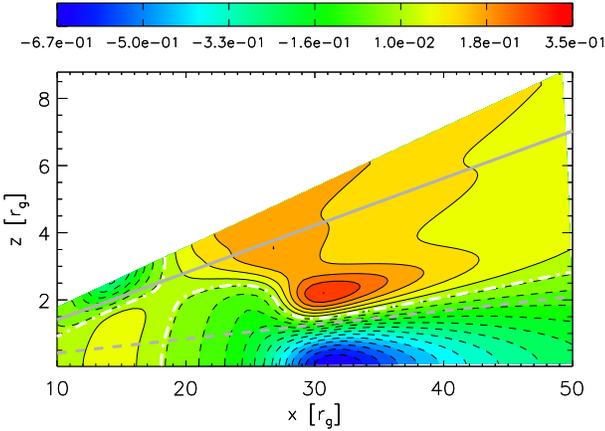}
 \end{center}
 \caption{Contour plot of the magnetic torque density (the integrand in 
equation~(\ref{eq:torque})) measured in units of $2.5 \times 10^{35}$~dyn, for 
configuration To~5 with $R_m^{\rm pol} = 207$ and $R_m^{\rm tor} = 2 \times 
10^{6}$. As usual, contours of negative torque are drawn with dashed lines and 
those for positive torque are drawn with solid lines. Regions with positive 
torque act in the sense of spinning up the star, and those with negative torque 
act in the sense of spinning it down. The white dot--dashed curves are zero 
lines.}
 \label{fig:integrand_cont}
\end{figure}

The contours for case (ii) are shown in Fig.~\ref{fig:integrand_cont}. The most 
striking feature of this is the region of positive spin-up which at high 
latitudes extends over all radii from about the corotation point outwards, 
which is contrary to what is usually thought. A second relevant feature is that 
there is a clear vertical structure, with the upper part being the spin-up 
region. Inwards of the corotation radius the structure is reversed, with the 
equatorial region contributing to the spin-up and the upper part to the 
spin-down. We recall that our $\Omega$ profile is basically Keplerian from the 
outer edge of the disc up until $r_1 = 16.67 \, r_{\rm g}$, inwards of which it 
is progressively changed so as to match with corotation at the inner edge of 
the disc.

The analysis of the torque density for the other three cases gives very similar 
results, with differences only in the details of the structure, e.g. the exact 
location of the boundary between the spin-up and spin-down regions. We have 
calculated the torque density also for the reference case of Paper~II, i.e. 
model number~$1$ in Table~\ref{tab:conf_SS}. We there again obtain spin-up 
regions also outside the corotation radius and see evidence of a vertical 
structure (in this case with two changes of sign as one moves from the corona 
towards the equatorial plane).

All of the cases show a strong relation between the magnetic torque density and 
the vertical structure of the toroidal component of the magnetic field. This 
can be seen more clearly if one compares the $B_\phi$ contour plot with the 
magnetic torque density plot. The sign of the torque density is opposite to 
that of the vertical derivative of the toroidal field $\partial_\theta 
B_\phi$, just as predicted by equation~(\ref{eq:torque}), when $T_B^r$ is 
smaller than $T_B^\theta$.

\begin{figure}
 \begin{center}
 \includegraphics[width=0.45\textwidth]{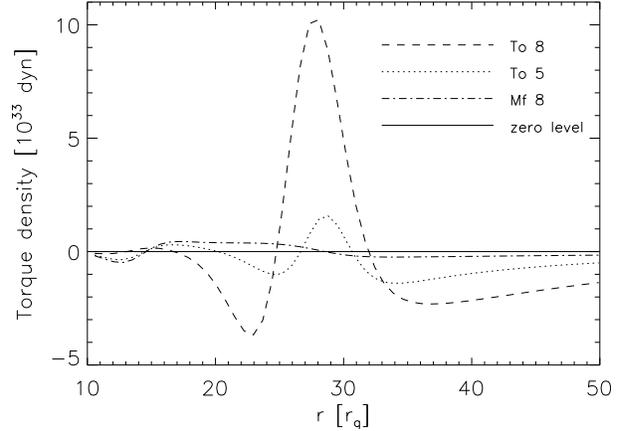}
 \end{center}
 \caption{Shell average of the magnetic torque density for configurations To~8, 
To~5 and Mf~8. The solid line marks the zero level, which is shown as a 
reference.}
 \label{fig:torque_avg}
\end{figure}

We have calculated a shell average of the magnetic torque density over the 
upper half of the disc to obtain an estimate of the total contribution to the 
torque from the disc at any given radial location. Results are shown in 
Fig.~\ref{fig:torque_avg} for cases (i) -- (iii)\footnote{The average 
for configuration To~12 is not shown because it is similar to that of To~8.}. 
This confirms that there are parts of the disc outwards of the corotation radius 
that spin the star up. Also, it shows that nothing special is happening for the 
magnetic torque at the corotation radius (at $18.8\,r_{\rm g}$). The toroidal 
magnetic field is zero there, but the magnetic torque depends only on the 
derivative terms.

\section{Comparison with our earlier models}\label{sec:CMP}
%
 In papers I and II, we considered the distortion of the magnetic field when 
the Shakura and Sunyaev disc velocity profile was used. Table~\ref{tab:conf_SS} 
lists some of the configurations described in Paper~II (which we will often 
refer to as PII in this subsection). For each of them, we have now calculated 
the torque as described in Section~\ref{sec:SUT} and the results of this are 
reported in Table~\ref{tab:torque_SS}.

\begin{table}
 \caption{Values of the magnetic Reynolds numbers for those configurations of 
Paper~II (PII) that are considered here. The last column reports some of the
configurations of the current paper which have the same or similar magnetic
Reynolds numbers.}
 \begin{center}
\begin{tabular}{cccccc}
\hline
Model      & $\eta_0$          & $v_0$   & $R_m^{\rm pol}$ & $R_m^{\rm tor}$ &
Close models  \\
(PII) & $( \rm{cm}^2 \, \rm{s}^{-1} )$ & $( \rm{cm} \,
\rm{s}^{-1} )$ & & & Current paper\\
\hline
1 & $\eta_0 = 10^{10}$        & $10^5$ & $41.4$  & $2\times 10^6$ &
Mf~1, Mf~5\\
2 & $\eta_0 = 4\times10^{10}$ & $10^5$ & $10.3$  & $5\times 10^5$ &
Fiducial, To~3\\
3 & $\eta_0 = 10^{11}$        & $10^5$ & $4.14$  & $2\times 10^5$ &
Mf~2~--~4\\
4 & $\eta_0 = 10^{12}$        & $10^5$ & $0.41$  & $2\times 10^4$ &
Mf~3, To~1, ~2\\
\hline
\end{tabular}
 \label{tab:conf_SS}
\end{center}
\end{table}
\begin{table}
 \caption{Values of the computed torque for the configurations of Paper~II, 
where the S\&S velocity profile was used. The quantities reported here are the 
same as in Tables~\ref{tab:torque_mf} and \ref{tab:torque_to}.}
 \begin{center}
\begin{tabular}{cccccc}
\hline
Model & $T_B$& $T_B^r$ & $T_{B-{\rm disc}}$ & $T_{\rm NS}$ \\
\hline
1 & $-21.6$ & $-17.0$ & $-20.2$ & $-19.6$ \\
2 & $-4.42$ & $-2.45$ & $-3.04$ & $-2.39$ \\
3 & $-1.06$ & $-0.05$ & $ 0.32$ & $ 0.97$ \\
4 & $-0.04$ & $ 0.07$ & $ 1.34$ & $ 1.99$ \\
\hline
\end{tabular}
 \label{tab:torque_SS}
\end{center}
\end{table}

There is no straightforward way of comparing between individual models in PII 
and the current paper. All of the models have a neutron star with the same mass 
and magnetic field, accreting at the same mass accretion rate and, in PII, even 
with the same $\alpha$-parameter\footnote{The mass accretion rate is fixed by 
our choice for the neutron-star mass, magnetic field and Alfv\'en radius. 
Varying $v_{0}$ corresponds to varying the effective $\alpha$, following the 
relation
 $$
\alpha = \left( \frac{v_0 }{4.6 \times 10^5} \right)^{5/4}.
$$ 
We were imprecise about this in our earlier papers, using a value of
$\alpha=0.10$ rather than $0.15$.}. The poloidal velocity fields are instead
very different. It seems best to consider the comparison in an overall sense.
Although the Reynolds numbers are useful indicators, we cannot here expect
models of the two types with the same pair of Reynolds numbers to give identical
results, because these numbers depend on the magnitude of the velocity fields
only at a specific location and do not take into account how the velocities
change with position. (For the turbulent diffusivity, we are using the same
profile as in PII.)

As regards the distortion of the magnetic field lines, in Section~\ref{sec:MFS}, 
we have seen that the distorted pattern is closely related to the behaviour of 
the fluid flow. Since in this paper we use a velocity field which is 
quite different from that in PII, we should not expect the poloidal magnetic 
field lines for the two cases to be similar. In PII, $v_r$ was always 
inward pointing, whereas now there is inflow at the higher latitudes but a 
backflow in the region of the mid-plane which is reflected in a bending 
backwards of the field lines in that region. In addition, the magnitude of the 
velocity field in the disc decreases strongly with increasing $\theta$ in the 
present calculations, whereas in PII it was constant. This strong decrease 
results in a weaker distortion for fixed $v_0$ and $\eta_0$.

For the toroidal direction, we use the same $\Omega$ profile in both papers. 
However $B_\phi$ and the magnetic torque also depend on the configuration of 
the poloidal magnetic field, so that even for configurations with the same 
$R_m^{\rm tor}$ (and $R_m^{\rm pol}$) we cannot expect to get the same results. 
In fact, the $B_\phi$ contours show an important difference, in that the large 
region of positive $B_\phi$ outside the corotation radius found in Paper~II, is 
no longer present here. The velocity field is the origin of this difference 
(nothing else has changed); however, it is not trivial to isolate the specific 
aspect which is causing this feature to disappear. Probably, it is because, with 
the new velocity profile, the distortions of the vertical component $B_\theta$ 
are smaller than in papers I and II (compare for example 
Fig.~\ref{fig:pol_eta0_4e10_Bpolavg} with fig. 4 of Paper~I).

As regards the magnetic torque, we can make a more quantitative comparison. For 
every model in PII, we can find a configuration in this work which has 
the same $R_m^{\rm pol}$ (or is very close to it). We then consider the ratio 
of the $R_m^{\rm tor}$ for the Paper~II model to that of this paper, and 
use this factor to scale the magnetic torque of the PII model. Results from 
doing this are reported in Table~\ref{tab:torque_cmp}. In general, the torques 
for PII are larger than what we find in the current calculations (in absolute 
value). This can be explained by the larger distortions found in PII. In 
addition to being larger, the distortions also follow a different pattern, and 
so the relative contributions of the two terms in the torque, as introduced in 
equation~(\ref{eq:torque_2terms}), are very different. Having an equatorial 
backflow in the velocity field distorts the magnetic field in such a way that 
the two parts of the magnetic torque often act in opposite directions and so 
the total torque is smaller than when $T^r_{\rm B-disc}$ is neglected. With the 
S\&S velocity field, where there is no flow reversal and $v_r$ is constant with 
$\theta$, the two terms tend to have the same sign, and therefore the total 
magnetic torque tends to be larger. However, the differences should 
progressively diminish for smaller and smaller $R_m^{\rm pol}$, because in the 
limit as $R_m^{\rm pol} \to 0$ the magnetic field tends to a dipole, regardless 
of the velocity field profile. This is nicely confirmed by the numbers in 
Table~\ref{tab:torque_cmp}, where the ratio of the torques becomes smaller as 
$R_m^{\rm pol}$ decreases.

\begin{table}
 \caption{Comparison between the torques found in the current work and in 
Paper~II (PII). The first two columns show which models are being compared. The 
third column gives the value of the poloidal magnetic Reynolds number for the 
PII model. The last column gives the ratio between the scaled PII torque and 
the torque of the equivalent model in the current paper.}
 \begin{center}
\begin{tabular}{cccc}
\hline
PII model & Current work model & $R_m^{\rm pol}$ PII & (Scaled $T_{B}^{\rm PII})
/ T_{B}$ \\
\hline
1 & Mf 5 & 41.4 & 4.6 \\
2 & To 3 & 10.3 & 5.1 \\
3 & Mf 3 & 4.14 & 3.3 \\
3 & Mf 4 & 4.14 & 3.7 \\
4 & To 1 & 0.41 & 1.3 \\
4 & To 2 & 0.41 & 1.3 \\
\hline
\end{tabular}
 \label{tab:torque_cmp}
\end{center}
\end{table}

\section{Conclusions}\label{sec:CON}
 This paper is part of a step-by-step analysis of the interaction between the 
magnetic field of a central rotating neutron star and an encircling accretion 
disc, focusing particularly on millisecond pulsars. Our strategy has been to 
start with a very simple model and then to progressively include additional 
features one at a time, so as to understand the effects of each of them. For 
the model presented here, we have considered a neutron star with a dipole 
magnetic field aligned with the rotation axis, surrounded by a disc which is 
truncated at the Alfv\'en radius and which has a coronal layer above and below 
it, treated as a boundary layer between the disc and its exterior. The external 
region is here taken to be vacuum and the magnetic field there is taken to be a 
perfect dipole. (Effects coming from interfacing on to a non-vacuum 
magnetosphere will be considered in subsequent work.) We have used the 
kinematic approximation, in which the velocity field of the matter is kept 
fixed, and have solved the induction equation for mean fields looking for a 
stationary configuration. We solve self-consistently for all of the magnetic 
field components, employing a velocity field having all three components 
different from zero, and do not neglect horizontal derivatives.

The code and the procedure used for this analysis have been described in detail 
in \citet{papI, papII}, where we used a specific 
three-dimensional extension of the \citet{SS73} velocity profile (i.e. a 
velocity field with three non-zero components, each depending at most on the 
two coordinates $r$ and $\theta$). Here, instead, we consider a more general 
prescription for the velocity field, as described in \citet{KK00}. The key 
features of this velocity field are (1) a fully three-dimensional axisymmetric 
profile and (2) the presence of a flow reversal (backflow) in a region close to 
the equatorial plane.

The magnetic configuration that we find is consistent with the conceptual 
picture described in papers I and II. As soon as magnetic field lines enter the 
corona, they start to be pushed inwards by the accreting matter. However, as one 
proceeds downwards in the disc, they start to be bent back again, as a 
consequence of the change in direction of the velocity field from inflow to 
backflow (see Fig.~\ref{fig:ref_Bpol}). When interaction with a non-vacuum 
magnetosphere is included, one expects to see the field lines above and below 
the disc being spread out with respect to the dipole configuration. It is 
interesting to compare our results with ones from those calculations which show 
this line inflation and also resolve the region inside the disc 
(e.g.~\citealp{MS97,ZF09,Rom11,ZF13}); it can be seen that 
our solutions for the magnetic field-line structure inside the disc are rather 
similar to those.

The toroidal component of the magnetic field in our present calculations 
shows a simpler profile than that in Paper~II (see Fig.~\ref{fig:ref_Bphi}). It 
is positive (dragged forwards with respect to the neutron star) inwards of 
the corotation point and negative (dragged backwards with respect to the 
neutron star) outwards of that, just as envisaged by \citet{W87} and 
\citet{C87} (i.e. as given by equation~\ref{eq:bp_an} when a dipole field is 
used for $B_z$). However, we find some crucial differences with respect to 
those models: (1) the relative strength of $B_\phi$ at the two extrema is 
reversed and, most importantly, (2) $B_\phi$ decreases with $r$ much more 
slowly, and this has consequences for the torque (see 
Fig.~\ref{fig:tor_eta0_4e10_B3cmp_avg}).

We find it convenient to introduce two magnetic Reynolds numbers to describe 
the magnetic field structure. One, $R_m^{\rm pol}$, is defined using a 
characteristic poloidal velocity and the other one, $R_m^{\rm tor}$, uses the 
Keplerian linear velocity at the corotation point. The former fully determines 
the structure of $B_{\rm pol}$ and the shape of $B_\phi$, while the latter 
fixes the magnitude of $B_\phi$. It is only when two configurations have the 
same pair of Reynolds numbers that one obtains the same results for all of the 
three components of the magnetic field. When the profile of the velocity or of 
the magnetic diffusivity is changed, the magnetic Reynolds numbers lose their 
predictive power and one should instead consider the magnetic distortion 
functions $D_m^{\rm pol}$ and $D_m^{\rm tor}$ (i.e. a two-dimensional 
generalization of the magnetic Reynolds numbers), meaning that configurations 
with the same $R_m$ but different $D_m$ give different results. As in Paper~II, 
we find that the magnetic distortion functions determine the position of the 
second peak in $B_\phi$; while the first peak is connected with $B_\theta 
\Delta\Omega$.

We have also made preliminary calculations of the total torque exerted on the 
neutron star. We split this into two parts: (1) the torque produced by the 
interaction between the magnetic field and the disc $T_{B-\rm disc}$ and (2) 
that produced by the accreting matter, leaving the disc at the inner edge and 
reaching the NS surface along magnetic field lines, $T_{B-{\rm acc}}$. The 
standard reference for calculating $T_{B-\rm disc}$ is \citet{GL79b}. Here we 
have modified that approach in two ways: (1) we drop some of the simplifying 
assumptions made in calculating the analytic expression for the torque (i.e. we 
retain both horizontal and vertical derivatives) and (2) for the magnetic field 
we use our numerical results, rather than the approximated profile of earlier 
models, although we recall our caveats about our present neglect of 
line-inflation and possible breaking of magnetic linkage beyond a certain 
radius in the disc. Regarding the dipolar boundary conditions: in Paper~I we 
showed that the shape of the field-lines inside the disc does not depend on the 
chosen boundary conditions. However, the magnitude of the field is affected, 
and the dilution introduced by line inflation would be expected to decrease the 
overall torque from the outer parts of the disc. (We will comment further on 
this below.) Regarding the breaking of magnetic linkage: we note that in the 
results of \cite{ZF09}, the point where magnetic linkage is lost comes at 
beyond twice their co-rotation radius. The situation which they are considering 
is rather different from ours (they are studying T Tauri stars), but if one 
transfers this result to our case, it would correspond to loss of linkage only 
at the very edge of the domain used in our plots, where the magnetic torque 
contribution is negligible anyway.

In the standard scenario developed in the 1980s and 1990s, and still widely 
used, the magnetic torque is taken to be proportional to $B_\phi$, which in 
turn is taken to be proportional to the difference between the local $\Omega$ 
of the disc matter and that of the central object, so that the disc inwards of 
the corotation radius gives a spin-up contribution, while the remaining part 
gives a spin-down. We find that this description, although rather reasonable, 
in fact fails to reproduce the behaviour of a simple two-dimensional disc 
model, even when the vertical averages of $B_\phi$ and the magnetic torque are 
considered (see Figs.~\ref{fig:tor_eta0_4e10_B3cmp_avg} and 
\ref{fig:torque_avg}). For any given configuration, there is a vertical 
structure, such that at any radius there are both spin-up and spin-down regions 
(at different heights in the disc, as shown in Fig.~\ref{fig:integrand_cont}) 
whose locations vary with the values of the magnetic Reynolds numbers. The sign 
of the torque mainly follows that of the vertical derivative of the toroidal 
field, $\partial_\theta B_\phi$. We find similar results also when calculating 
the torque for the models of papers I and II, with an S\&S velocity profile.

Within our present model, the torque contributions from the inner part of 
the disc and from the accreting matter inwards of its inner edge are both 
positive (acting in the sense of spinning up the neutron star) while that from 
the rest of the disc can be either positive or negative. We have shown that the 
behaviour of the latter is complicated, and needs to be calculated with care 
rather than relying on the standard simplified prescriptions. However, for our 
present restricted model, we find that the net contribution from this region 
would still be negative (acting in the direction of spin-down), except in 
extreme cases, and that it would generally dominate over the positive 
contributions. This cannot be right for the spinning-up of millisecond pulsars, 
and so we infer that the action of field-line inflation and reconnections in 
weakening the negative torque from the outer parts of the disc probably plays a 
crucial role in allowing the overall spin-up.

In this paper, we have extended our previous work, where we used the S\&S 
velocity profile, by using instead the more sophisticated profile given by 
K\&K, and we have also made calculations for the torque exerted by the disc on 
the neutron star. The next step of our programme of work will be to replace the 
present vacuum boundary conditions by a consistent treatment of the interface 
with a non-vacuum force-free magnetosphere.

\section*{Acknowledgements}
 LN has been supported by a Chinese Academy of Sciences fellowship for young 
international scientists (Grant Number 2010Y2JB12). The Chinese Academy of 
Sciences and the National Astronomical Observatory of China (NAOC) of CAS have 
supported this work within the Silk Road Project (CAS Grant Number 2009S1-5) 
and it has also been partially supported by the Polish NCN grant NN203381436.

\bsp

\label{lastpage}

\end{document}